\documentclass[]{aa}

\usepackage{amsmath}
\usepackage{graphicx}
\usepackage{natbib}
\usepackage[varg]{txfonts}
\usepackage[UKenglish]{babel}
\usepackage[latin1]{inputenc}
\usepackage{multirow}
\bibpunct{(}{)}{;}{a}{}{,}
\usepackage{indentfirst} 
\usepackage{newlfont}                                                      
\usepackage{amssymb} 
\usepackage{latexsym}
\usepackage{longtable}
\usepackage{supertabular}
\usepackage[toc,page]{appendix}

\def\today{\ifcase\month\or January\or February\or March\or April\or May\or
      June\or July\or August\or September\or October\or November\or December\fi
      \space\number\day, \number\year}

\begin{document}

\title{MiNDSTEp differential photometry of the gravitationally lensed quasars WFI\,2033-4723 and HE\,0047-1756: Microlensing and a new time delay
\thanks{Based on data collected by MiNDSTEp with the Danish 1.54 m
    telescope at the ESO La Silla observatory.}}

  \author{\tiny{
       E. Giannini \inst{1}
  \and R. W. Schmidt\inst{1}
  \and J. Wambsganss \inst{1}
  \and K.~Alsubai\inst{2}
  \and J.~M.~Andersen \inst{3,4}
  \and T.~Anguita\inst{5,39}
  \and V.~Bozza\inst{7,8}
  \and D.~M.~Bramich\inst{2}
  \and P.~Browne\inst{10}
  \and S.~Calchi Novati\inst{7,13,37}\thanks{Sagan visiting fellow}
  \and Y.~Damerdji \inst{14}
  \and C.~Diehl \inst{1,15}
  \and P.~Dodds\inst{10}
  \and M.~Dominik\inst{10} \thanks{Royal Society University Research Fellow}
  \and A.~Elyiv\inst{14,18,38}
  \and X.~Fang \inst{36}
  \and R.~Figuera~Jaimes \inst{10,9}
  \and F.~Finet\inst{17}
  \and T.~Gerner\inst{1,6}
  \and S.~Gu \inst{36,42}
  \and  S.~Hardis\inst{20}
  \and K.~Harps\o e\inst{20,4}
  \and T.~C.~Hinse\inst{22,41}
  \and A.~Hornstrup\inst{23}
  \and M.~Hundertmark\inst{20,4,10,16}
  \and J.~Jessen-Hansen\inst{19}
  \and U.~G.~J\o rgensen\inst{20,4}
  \and D.~Juncher \inst{20,4}
  \and N.~Kains\inst{10}
  \and E.~Kerins\inst{24}
  \and  H.~Korhonen \inst{20,4}
  \and C.~Liebig\inst{10,1}
  \and  M.~N.~Lund \inst{19}
  \and  M.~S.~Lundkvist \inst{19}
  \and  G.~Maier\inst{1,6}
  \and  L.~Mancini\inst{6,7}
  \and G.~Masi\inst{26}
  \and M.~Mathiasen\inst{20}
  \and M.~Penny\inst{21,24}
  \and S.~Proft\inst{1}
  \and  M.~Rabus \inst{27,6}
  \and S.~Rahvar\inst{28,29}
  \and D.~Ricci\inst{30,31,40}
  \and G.~Scarpetta\inst{7,8}
  \and  K.~Sahu\inst{32}
  \and S.~Sch\"afer\inst{16}
  \and F.~Sch\"onebeck\inst{1}
  \and  J.~Skottfelt\inst{11,4}
  \and C.~Snodgrass\inst{33,34}
  \and J.~Southworth \inst{35}
   \and J.~Surdej\inst{14} \thanks{also Directeur de Recherche honoraire du FRS-FNRS}
    \and J.~Tregloan-Reed \inst{12,35}
  \and C.~Vilela \inst{35}
  \and  O.~Wertz\inst{14}
  \and F.~Zimmer\inst{1}
}}

\institute{
    \tiny{Astronomisches Rechen-Institut, Zentrum f\"ur Astronomie, Universit\"at Heidelberg, M\"onchhofstra\ss e 12-14, 69120 Heidelberg, Germany \
\email{emanuela@ari.uni-heidelberg.de}
\and 
    Qatar Environment and Energy Research Institute (QEERI), HBKU, Qatar Foundation, Doha, Qatar
\and 
    Department of Astronomy, Boston University, 725 Commonwealth Avenue, Boston, MA 02215, USA 
    \and  
    Niels Bohr Institute \& Centre for Star and Planet Formation, University of Copenhagen {\O}ster Voldgade 5, DK-1350 Copenhagen, Denmark
    \and
    Departamento de Ciencias F\'isicas, Universidad Andres Bello, Avenida Rep\'ublica 220, Santiago, Chile
    \and
    Max-Planck-Institut f\"ur Astronomie, K\"onigstuhl 17, 69117 Heidelberg, Germany
    \and
    Dipartimento di Fisica ''E. R. Caianiello'', Universit\`a di Salerno, Via Giovanni Paolo II 132, 84084-Fisciano (SA), Italy
    \and
    Istituto Nazionale di Fisica Nucleare, Sezione di Napoli, Italy
    \and
    European Southern Observatory, Karl-Schwarzschild-Stra\ss e 2, 85748 Garching bei M\"{u}nchen, Germany
    \and 
    SUPA, University of St~Andrews, School of Physics \& Astronomy, North Haugh, St~Andrews, KY16 9SS, United Kingdom
    \and
    Centre for Electronic Imaging, Dept. of Physical Sciences, The Open University, Milton Keynes MK7 6AA, UK
    \and 
    NASA Ames Research Center, Moffett Field CA 94035, USA.
    \and
    Istituto Internazionale per gli Alti Studi Scientifici (IIASS), Vietri Sul Mare (SA), Italy
    \and
    Institut d'Astrophysique et de G\'eophysique, Universit\'e de Li\`ege, All\'ee du 6 Ao\^ut, B\^at. B5c, 4000 Li\`ege, Belgium
    \and
    Hamburger Sternwarte, Universit\"at Hamburg, Gojenbergsweg 112, 21029 Hamburg, Germany
    \and
    Institut f\"ur Astrophysik, Georg-August-Universit\"at G\"ottingen, Friedrich-Hund-Platz 1, 37077 G\"ottingen, Germany
    \and
   Subaru Telescope, National Astronomical Observatory of Japan, 650 North Aohoku Place, Hilo, HI 96720
    \and
    Main Astronomical Observatory, Academy of Sciences of Ukraine, Zabolotnoho 27, 03680 Kyiv, Ukraine
  \and
    Stellar Astrophysics Centre, Department of Physics \& Astronomy, Aarhus University, Ny Munkegade 120, DK-8000 Aarhus C, Denmark
  \and
    Dark Cosmology Centre, Niels Bohr Institute, University of Copenhagen, Juliane Maries vej 30, DK-2100 Copenhagen {\O}, Denmark
  \and
    Department of Astronomy, Ohio State University, 140 W. 18th Ave., Columbus, OH 43210, USA
  \and
    Korea Astronomy \& Space Science Institute (KASI), 305-348 Daejeon, Republic of Korea
  \and
    National Space Institute, Technical University of Denmark, 2800 Lyngby, Denmark
    \and 
  Jodrell Bank Centre for Astrophyics, University of Manchester, United Kingdom
  \and
  Finnish Centre for Astronomy with ESO (FINCA), University of Turku, V{\"a}is{\"a}l{\"a}ntie 20, FI-21500 Piikki{\"o}, Finland
  \and
  Bellatrix Astronomical Observatory, Center for Backyard Astrophysics, Ceccano (FR), Italy
  \and
  Centro de Astro-Ingenier\'ia, Instituto de Astrof\'isica, Facultad de F\'isica, Pontificia Universidad Cat\'olica de Chile, Av. Vicu\~na Mackenna 4860, 7820436 Macul, Santiago, Chile
  \and
  Physics Department, Sharif University of Technology, Tehran, Iran
  \and
  Perimeter Institute for Theoretical Physics, 31 Caroline Street North, Waterloo, Ontario N2L 2Y5, Canada
  \and
   Observatorio Astron\'omico Nacional, Instituto de Astronom\'ia -- Universidad Nacional Aut\'onoma de M\'exico, Ap. P. 877, Ensenada, BC 22860, Mexico
  \and
  Instituto de Astrof\'isica de Canarias, E-38205 La Laguna, Tenerife, Spain
  \and
  Space Telescope Science Institute (STScI), United States of America
  \and
 Planetary and Space Sciences, Department of Physical Sciences, The Open University, Milton Keynes, MK7 6AA, UK
    \and 
  Max-Planck-Institut f\"ur Sonnensystemforschung, Justus-von-Liebig-Weg 3, 37077 G\"ottingen, Germany
  \and 
 Astrophysics Group, Keele University, Newcastle-under Lyme, ST5 5BG, United Kingdom
  \and
 Key Laboratory for the Structure and Evolution of Celestial Objects, Chinese Academy of Sciences, Kunming 650011, China
  \and
NASA Exoplanet Science Institute, MS 100-22, California Institute of Technology, Pasadena CA 91125
\and
 Dipartimento di Fisica e Astronomia, Universit\`a di Bologna, viale Berti Pichat 6/2, 40127  Bologna, Italy
\and
 Millennium Institute of Astrophysics, Chile
\and
 Universidad de La Laguna, Departmento de Astrof\'isica, E-38206 La Laguna, Tenerife, Spain
\and
Armagh Observatory, College Hill, BT61 9DG Armagh, United Kingdom
\and
Yunnan Observatories, Chinese Academy of Sciences, Kunming 650216, China
}}

\date{Draft \today} 
\authorrunning{E.Giannini et al.}
\titlerunning{MiNDSTEp differential photometry of the lensed quasars WFI\,2033-4723 and HE\,0047-1756}

 \abstract
 {} {We present V and R photometry of the gravitationally lensed quasars WFI\,2033-4723 and HE\,0047-1756.
 The data were taken by the MiNDSTEp 
 collaboration with the 1.54 m Danish
 telescope at the ESO La Silla observatory from 2008 to 2012.} {Differential photometry has been carried out using the image subtraction method as implemented in the HOTPAnTS package,
 additionally using GALFIT
 for quasar photometry.}{The quasar WFI\,2033-4723 showed brightness variations of order 0.5 mag in V and R during the campaign.
 The two lensed components of quasar  HE\,0047-1756 varied by 0.2-0.3 mag within five years. We provide, for the first time, an estimate of the time delay
 of component B with respect to A of $\Delta t= (7.6\pm1.8)$ days for this object.
 We also find evidence for a secular evolution of the magnitude difference between components A and B in both filters, which we explain as due to a long-duration
 microlensing event. 
 Finally we find that both quasars WFI\,2033-4723 and HE\,0047-1756 become bluer when brighter, which is consistent with previous studies.}{}

\maketitle

\section{Introduction}

Quasar microlensing is caused by compact objects along the line of sight towards quasars, which are multiply imaged by foreground lensing galaxies
(\cite{1979Natur.282..561C}, \cite{1981ApJ...243..140G}, \cite{1981ApJ...244..736Y}).
The perturbative effect on the quasar images
consists of brightness variations up to a magnitude over timescales of weeks to years.
Multiply imaged quasars are particularly suitable for isolating microlensing variations, which rise in an uncorrelated fashion between the images,
in contrast to the intrinsic quasar fluctuations, which appear in all quasar images after a certain time delay. Quasar microlensing can be used as a method to study the structure of quasars, since
the amplification of the microlensing signal depends on the size of the quasar emitting region. It also works as a probe for the existence of compact objects between
the observer and the quasar and for their mass distribution. Moreover, the measurement of time delays constitutes an indirect measurement of the cosmological constant
$H_0$ (\cite{1964MNRAS.128..307R}).\\

Here we present the multi-band photometry of two lensed quasars, WFI\,2033-4723 and HE\,0047-1756,
in the V and R spectral bands, observed with the 1.54 m Danish telescope at the ESO La Silla observatory (Chile), in the framework of the MiNDSTEp quasar monitoring campaign from 2008
to 2012.
We applied the \cite{2000A&AS..144..363A} image subtraction method (see also \cite{1998ApJ...503..325A}), as implemented in the HOTPAnTS
subtraction package by A. Becker (Becker et al. (2004)), and then carried out difference image photometry. The main advantage of this 
approach is that photometry on difference images does not require us to model the foreground lens galaxy, since it is removed after subtraction. Below
we summarize the main properties of the observed quasars (Section ~\ref{sect:obj}).
In Section ~\ref{sect:obs} we present the observations of the quasars. Section ~\ref{sect:da} treats the image subtraction method at length. The light curves
of WFI\,2033-4723 and HE\,0047-1756 are shown in Section ~\ref{sect:res}, which also includes a measurement of the time delay between the components of
HE\,0047-1756. We discuss our results in Section ~\ref{sect:summ}.
 
\section{Short notes on WFI\,2033-4723 and HE\,0047-1756}

\label{sect:obj}
\subsection{WFI\,2033-4723}
The quadruply imaged quasar WFI\,2033-4723 , see Fig.~\ref{fig:wfi2033_stamps}, was discovered by \cite{2004AJ....127.2617M}; the four lensed images at redshift $z_Q=1.66$ show a maximum angular separation of $2.53''$.
\cite{2006A&A...451..759E} found that the lens galaxy spectrum is consistent with an elliptical or S0 template at redshift $z_L=0.661\pm0.001$.
\cite{2008A&A...488..481V} determined the time delays as $\Delta t_{B-C}= 62.6^{+4.1}_{-2.3}$ days and $\Delta t_{B-A}=35.5\pm1.4$ days between components
C and B, and A and B, respectively, where A indicates the combination of images A1 and A2. Since A1 and A2 are predicted to have a negligible relative time delay,
they are treated as a blend by Vuissoz et al. (2008).

\subsection{HE\,0047-1756}
The quasar HE\,0047-1756, see Fig.~\ref{fig:he0047_stamps}, was discovered in the ESO Hamburg quasar survey
(\cite{1996A&AS..115..227W}, \cite{1996A&AS..115..235R}, \cite{2000A&A...358...77W}).
\cite{2004A&A...419L..31W}
found that the quasar is in fact a lensed system with two observable images that are separated by $\Delta\theta=1.44''$ and these authors estimated the quasar redshift
at $z_Q=1.67$. The lensing galaxy, also discovered by \cite{2004A&A...419L..31W}, using the Magellan 6.5 m telescopes on Cerro Las Campanas, is at redshift 
$z_L=0.408$, according to \cite{2006ApJ...641...70O} (see also \cite{2006A&A...451..759E}). The lens galaxy spectrum matches well with an elliptical galaxy template
(\cite{2006ApJ...641...70O}, \cite{2006A&A...451..759E} ). The time delay has not been measured yet.

\section{Observations}
\label{sect:obs}

\subsection{Data aquisition}

\begin{figure}[htbp]
\begin{center}
\includegraphics[scale=1,width=\columnwidth]{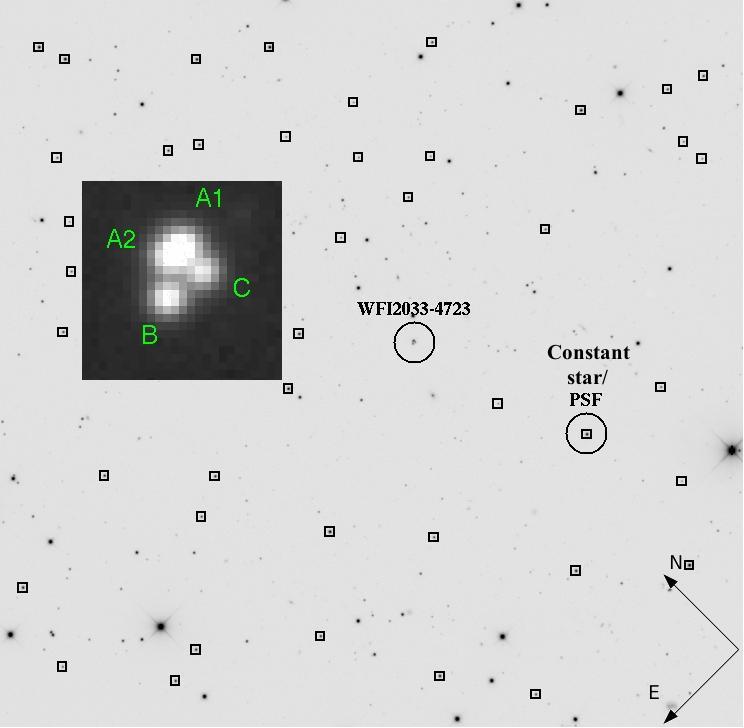}
\end{center}
\caption{V-band image of WFI\,2033-4723 obtained by stacking the 14 best seeing and
sky background images (V-band template image). The quasar and star, which we use both as a constant reference and PSF model, are 
labelled. The four lensed QSO components are
enlarged in the darker box. The field size is $9.7$ arcmin 
$\times$ $9.5$ arcmin.   
The stamps used for the kernel computation are defined as $17$-pixel squares (see Sect. 4).}
\label{fig:wfi2033_stamps}
\end{figure}

\begin{figure}[htbp]
\begin{center}
\includegraphics[scale=1,width=\columnwidth]{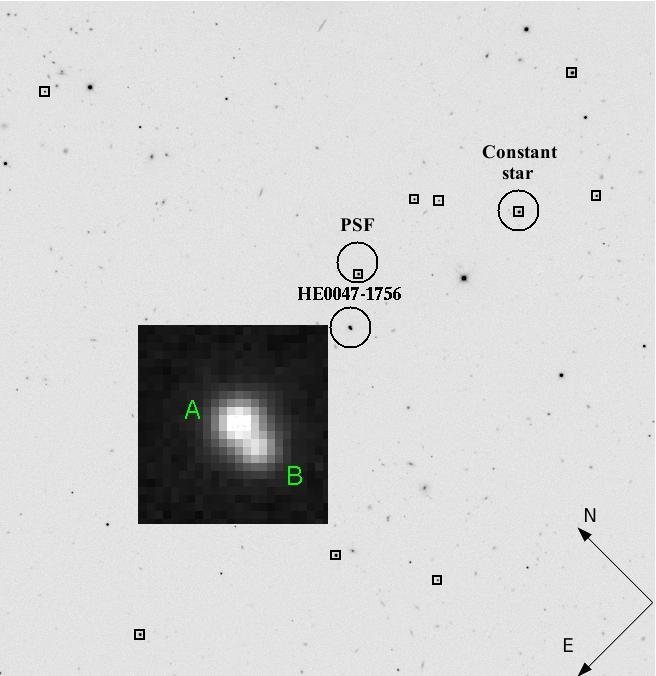}
\end{center}
\caption{V-band image of HE\,0047-1756 obtained by stacking the 10 best seeing and
sky background images (V-band template image). The quasar and stars, which we use as a constant reference and PSF model, are labelled.
The double quasar is enlarged in the darker box.
The field size is $8.5$ arcmin 
$\times$ $8.8$ arcmin.
The stamps used for the kernel computation are defined as $17$-pixel squares (see Sect. 4).}
\label{fig:he0047_stamps}
\end{figure}

The observations of the lensed quasars WFI\,2033-4723 and HE\,0047-1756 were obtained
in the $V$ and $R$ bands  with the 1.54\,m Danish
telescope at the ESO La Silla observatory, Chile, in the framework of the MiNDSTEp (Microlensing network for the detection of small terrestrial exoplanets;
\cite{2010AN....331..671D}) quasar monitoring campaign. 
We report here on observations collected during five observing seasons from 2008 to 2012.
Both quasars were observed within the following temporal intervals: from June 5 to October 4 2008, June 19 to September 18 2009, May 9 to August 21 2010, June 11 to August 31 2011, and July 16 to September 16 2012. During these periods we observed the quasars every
two days weather permitting. 
The $V$ and $R$ filters provide photometry in the Bessel system. The quasars were observed on average three times
per night. WFI\,2033-4723 was observed 
with an exposure time of $180$ s in $V$ and $R$, except
for a small number of images, with longer exposures
of $300$ s and $600$ s, at the start of 2008. Exposure times for HE\,0047-1756 
vary from $180$ s in both filters during the first two seasons to $240$ s
 in V and $210$ s in R during the last three years. A few images in both bands
at the start of the first season were taken with exposure times
of $300$ s and $600$ s. The median seeing of the observations is $\approx 1.3$ arcsec, taking both filters and all data into account.
The observations were made with the
%
Danish Faint Object Spectrograph and Camera (DFOSC)
imager with a pixel scale of 0.39 arcsec. 
The full field
of view (FOV) was 13.7 arcmin $\times$ 13.7 arcmin. We subtracted constant bias values 
estimated from the overscan regions to carry out bias correction and used dome flat-field frames for flat-fielding.
Sections of the FOV centred on the quasars are shown in Figs.~\ref{fig:wfi2033_stamps} and~\ref{fig:he0047_stamps}.

\section{The method}
\label{sect:da}

\begin{figure}[htbp]
\begin{center}
\includegraphics[width=\columnwidth]{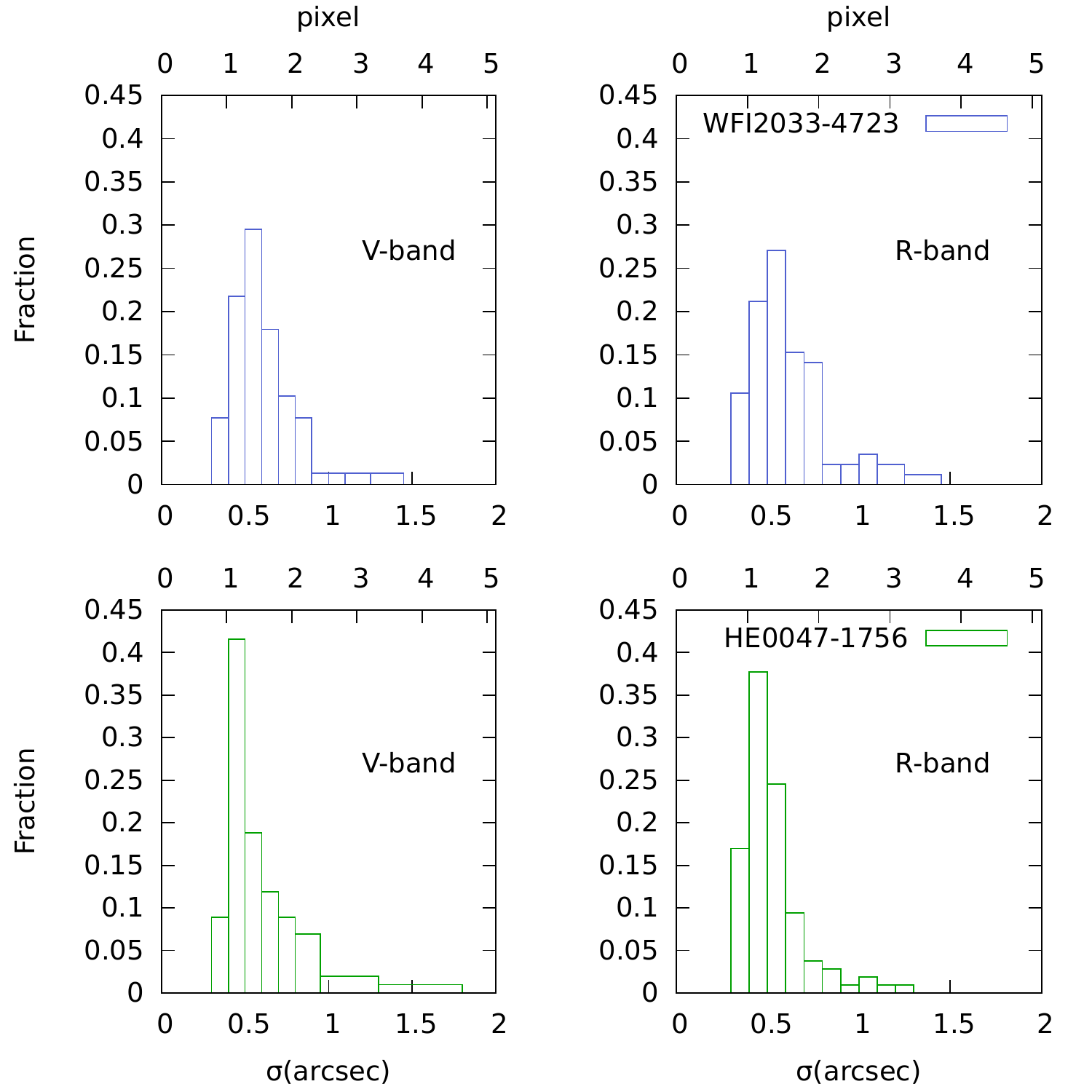}
\end{center}
\caption{FWHM distributions of a nearby star of WFI\,2033-4723 (top) and HE\,0047-1756 (bottom) in filters V (left) and R (right) in terms
of the corresponding Gaussian $\sigma$.}
\label{fig:sigmadistr}
\end{figure}

As was shown impressively in the case of the Huchra lens
(\cite{2000ApJ...529...88W}, \cite{2006AcA....56..293U}), arguably the best way to
carry out photometry for a quasar lens is the difference image
analysis technique (DIA) proposed by \cite{1998ApJ...503..325A} and \cite{2000A&AS..144..363A}. The basic
idea is to use a high signal-to-noise (S/N) template image, with good seeing and low sky
background, which is
subtracted from every other target frame in the data set. Since the lensing galaxy is not expected to vary, this approach simplifies the
photometry of the quasars. As the contribution from the galaxy 
is removed in the subtracted images, we are spared the drawbacks of modelling the galaxy
light distribution. 
Each pair of images needs to be astrometrically
and photometrically aligned before subtraction. Relative photometry can be carried out upon subtraction. This is achieved by building a model for the quasar images with
a blend of known point spread functions (PSFs) according to Hubble space telescope (HST) astrometry of the quasar.\\
In detail, the idea from \cite{1998ApJ...503..325A} is to find the best-fit, spatially non-varying convolution kernel, which degrades the template PSF into that of the target frame in addition to matching atmospheric extinction and
exposure time. These authors showed that, by decomposing the kernel as a linear combination of N basis functions, the problem can be reduced to determining a finite number of
kernel coefficients. The latter are found by solving a linear system of equations containing
various moments of the two images in input. The analytical convolution 
kernel consists of a sum of several Gaussians,
which are multiplied by polynomials modelling the possible asymmetry
of the kernel. The Gaussian widths depend on the relative sizes of the PSF in the template and target frame. \cite{2000A&AS..144..363A} extended this idea
to a kernel that varies across the chip. Assuming that the amplitudes of the kernel components are polynomial
functions of the pixel coordinates of order n, the number of kernel coefficients becomes $\frac{1}{2}(n+1)(n+2)$ larger
than in the constant kernel problem. 

\subsection{Image alignment}

Before subtraction, all images need to be astrometrically aligned to a reference frame. 
After choosing a very good seeing and low sky background reference image of a given source,
all the other images of the
same source are registered onto the reference coordinate grid, whether or not they share the same photometric band with the reference image.
This is carried out using the ISIS%
\footnote{http://www2.iap.fr/users/alard/package.html} package by
C. Alard (\cite{1998ApJ...503..325A}, \cite{2000A&AS..144..363A}). The astrometric alignment routine of this package
efficiently identifies reference objects (stars) in the field and
carries out a two-dimensional polynomial mapping to the reference image.
We choose a polynomial transformation of order 2 to remove the shifts
and small rotations between the images. In the case of WFI\,2033-4723, the average residuals corresponding
to the astrometric transform along the x- and y-axes are of order 0.1 pixel and the mapping
is computed using on average  $\approx 275$ objects. The astrometric transforms corresponding to
HE\,0047-1756 are characterized by average residuals of order 0.2 pixel and are obtained taking on average $\approx 220$ objects into account.
Image resampling is performed using bicubic splines.  Before calling the
alignment routine, bad columns are replaced with the appropriate
median of surrounding pixels. All images of a
given night for the whole data set are trimmed at the edges to
contain the same region and median combined to improve the signal-to-noise ratio
and correct for cosmic rays. 
After discarding images that have a high sky background or are disturbed by moonlight, clouds, bad tracking, and bad columns,
we finally used 85 nights for WFI\,2033-4723 and 108 nights for HE\,0047-1756 over five years. V- and R-band images are not always
both available for any given night.
\subsection{Image subtraction}

\begin{figure*}[htbp]
\begin{center}
\includegraphics[scale=1,width=0.7\textwidth]{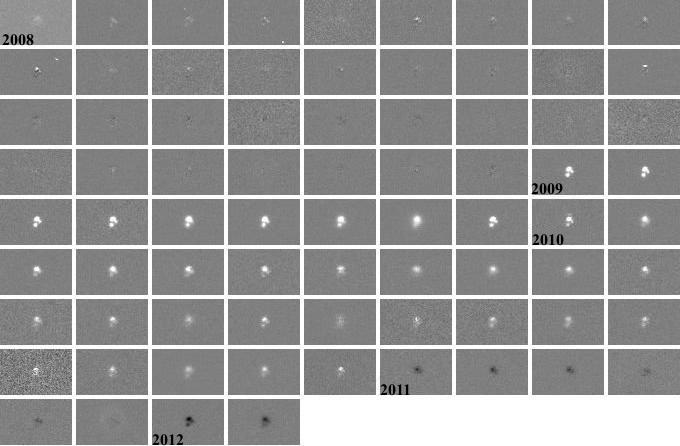}
\end{center}
\caption{V-band difference images of WFI\,2033-4723 from 2008 to 2012. The corresponding dates are listed in Table~\ref{datawv}.}
\label{fig:wfi2033V_sub_seq}
\end{figure*}
Image subtraction is carried out with the HOTPAnTS%
\footnote{http://www.astro.washington.edu/users/becker/hotpants.html}
software by A. Becker, which is an enhanced and
modified version of ISIS and the \cite{2000A&AS..144..363A} method. This software
is given the template image and the target frames to be processed. 
In creating the template images we stack the frames with the best seeing and
sky background at our disposal. In the case of WFI\,2033-4723 we compute the median stack of 14
images, both in V and R.
A similar procedure is carried out in building the template images for HE\,0047-1756,
for which we were able to combine 10 frames in V and 14 frames in R. 
The HOTPAnTS software divides the frames horizontally and vertically into square regions, within which several stamps, centred on individual stars, are chosen. The software is also given the 
list of stars that act as stamps. A kernel solution is derived for each stamp.
The kernel sum is used as a first metric
to sigma-clip bad stamps. Briefly, the photometric scaling between two images is
the sum of the convolution kernel. This can be used to discard variable stars, which are not
suitable for determining the photometric alignment between the images, by sigma-clipping outlier stamps from the
distribution of the kernel sums. It is useful to have multiple stars in a particular image region in case
any objects are sigma-clipped. In fact at this stage another metric is used to discard bad stamps. After convolution and subtraction,
the mean of the distribution of pixel residuals divided by the estimated pixel variance across each stamp provides
an additional figure of merit to sigma-clip stamps out and replace them with neighbouring stars.
The constraints on the convolution kernel for each stamp then
allow for the fitting of the coordinate dependent amplitudes of the kernel components (modelled as polynomial
functions). \\

For the analytical kernel we choose three Gaussians. With the aim of modelling the wings and the asymmetry of the kernel, the Gaussians are modified by 
multiplicative polynomials of orders 4, 3, and
2, respectively. In general we choose
a narrow Gaussian that varies to order 4, a central wider Gaussian that varies 
at order 3, and a broad Gaussian that varies at order 2 in the kernel space coordinates. 
The values of the Gaussian widths $\sigma$ depend on the seeing range of the templates and target frames
fed to HOTPAnTS. For the target frames we compute the seeing distribution of a star in the quasar neighbourhoods in the V and R bands.
We obtain
the corresponding $\sigma$ distribution, as shown in Fig. ~\ref{fig:sigmadistr},
via the relation between the FWHM and the $\sigma$ of a Gaussian profile. The triplet (0.8,1.4,2.3) in pixel units is a good choice 
to build the kernel basis which, convolved with the typical template $\sigma$ of 1 pixel, reproduces the typical target frame $\sigma$ values.
We adopt these values as our Gaussian widths. The polynomials modelling the spatial variation of the
kernel amplitudes are allowed to vary at spatial order 2 in the image space coordinates. 

The number of stamps differs from case to case since it depends on the star distribution in
the field.
The field characterizing the frames
of WFI\,2033-4723 is rich with stars, while that corresponding to HE\,0047-1756 is very sparse.
The convolution kernel of WFI\,2033-4723 is derived by taking into account on average 27 stamps in V and 26
in R; on average 8 stamps in V and 10 in R determine the convolution kernel for HE\,0047-1756.
The size of the stamps, 17$\times$17 pixels, is chosen such that it contains the whole star flux profiles. 
In summary, for each image, the code chooses stamps among the 
stars, works out
the convolution kernel and carries out the subtraction. Figures ~\ref{fig:wfi2033_stamps} and~\ref{fig:he0047_stamps} show the squares defining the stamps that HOTPAnTS
selects across all observing seasons in the filters V and R for the two systems.\\

Necessary inputs for the HOTPAnTS software are the target frame and template gain (G) and readout noise (RON). Throughout the 2008-2011 seasons
the instrumental gain G was $0.76 e^-/ADU$ and the readout noise RON was $3.21 e^-$. The G and RON values changed in 2012 as a result of an upgrade of the DFOSC detector between 2011 and 2012. The values that are valid for 2012 are $G=0.24 e^-/ADU$ and $RON=5.28 e^-$.
The gain and RON have to be adjusted
appropriately before feeding them to the HOTPAnTS software. Since both the target frames and the templates
are in general stacked images, we need to define an effective G and RON by using standard variance propagation.
Another adjustable set of parameters is the model
for the sky background, which we selected to be an additive constant.
The output by HOTPAnTS is the difference image with the seeing of the current target frame and the photometric scale of the template,
and the corresponding noise map. Figure ~\ref{fig:wfi2033V_sub_seq} shows the sequence of difference images for WFI\,2033-4723 in filter V.

\subsection{Photometry}

Photometry on the difference images is carried out using the GALFIT
software (version 2.0.3, \cite{2002AJ....124..266P}) modified to allow the fitting of several PSFs with
fixed relative separations and linear fluxes. We use it to analyse the lensed
quasars in the original images and in the
difference images. This is performed as follows:
\begin{enumerate}
\item A nearby star is chosen as a PSF model
(the chosen stars are labelled as PSF in Figs.~\ref{fig:wfi2033_stamps} and~\ref{fig:he0047_stamps}). We use only one star 
since we noticed a remarkable variation of the PSF through the field and decided to select the closest bright, isolated, and not saturated
star in the neighbourhoods. 
GALFIT normalizes the star so that
variability is not an issue.
The PSF is built by
extracting a 17$\times$17 pixel box surrounding the star; the PSF is sky-subtracted
and centred at pixel (9,9) according to the GALFIT manual.
\item Using the PSF model, the
  quasar positions are determined from the original stacked images and the templates with GALFIT by keeping the quasar
  fluxes and the position of only one of the lensed quasar images as 
  free parameters. The sky background values at the quasar position are fixed
 to median values estimated from 51$\times$51 pixel empty regions
 nearby the quasar. The positions of the remaining lensed quasar components relative to the free quasar component are
  kept fixed at the values shown in Table ~\ref{tab:coords}, which are obtained from the CASTLES \footnote{http://www.cfa.harvard.edu/castles/} web page
(C.S. Kochanek, E.E. Falco, C. Impey, J. Lehar, B. McLeod, H.-W. Rix), as determined using Hubble Space Telescope data.

\begin{table}[htbp]
\large
  \caption{Hubble Space Telescope relative astrometry of WFI\,2033-4723 and HE\,0047-1756 images, obtained from the CASTLES webpage.}
\label{tab:coords}
\centering
\scalebox{0.67}{
\hskip-0.cm\begin{tabular}{llp{2.cm}p{2.5cm}p{2.cm}lp{2.cm}}
\hline\hline
\multirow{3}*{WFI2033}   && A1 & A2 & B &C\\
\hline
&RA('') & $2.196\pm0.003$ & $-1.482\pm0.003$ & $0$ & $-2.114\pm0.003$\\
 &DEC('') & $1.261\pm0.003$ & $1.376\pm0.003$ & $0$ & $-0.277\pm0.003$\\
\hline
\multirow{3}*{HE0047}  && A &B &  & \\
\hline
 &RA('') & $0$ & $0.232\pm0.003$ & & \\
 &DEC('') & $0$ & $-1.408\pm0.003$& &\\
\hline
\end{tabular}
}
\end{table}

   Since the original stacked images and template images in general are built from a number of single exposures
with different exposure times, we let GALFIT build the appropriate sigma image 
by providing it with the equivalent GAIN and RDNOISE of the frames, since GALFIT is only able to
compute the noise image for a stack of N images with identical gain, readout noise, and exposure time. In our data we cannot significantly detect the
  lensing galaxies. Our positions are therefore minimally
  affected by the presence of the $V\approx 21$ mag (WFI\,2033-4723) and  $V\approx 22.5$ mag (HE\,0047-1756) lensing galaxy.
\item Keeping the nightly lensed quasar positions obtained above fixed, we determine
  the fluxes at the position of the quasar images in the difference images. For this, the GALFIT software is allowed to
  fit negative fluxes as well. The result of this procedure are
  difference fluxes between the epoch considered and the template image.
The output noise map from HOTPAnTS is used for the difference image photometry with GALFIT.\\

The robustness of the method just described has also been tested by computing the light curves of
the four components of quasar HE0435-1223 (\cite{2000A&A...358...77W}, \cite{2002A&A...395...17W}), which was
observed by MiNDSTEp and already published by \cite{2011A&A...528A..42R}, and comparing
them with the results in the R band by \cite{2011A&A...536A..53C}.
The four light curves, obtained with two different telescopes 
and two different methods, show an average weighted root mean square deviation of $rms=1.36\sigma$. 
A similar test has been carried out by computing with this method the light curves of quasar UM673 (\cite{1982ApJ...261..412M}, \cite{1987Natur.329..695S},
\cite{1988A&A...198...49S}), \cite{1992ApJ...389...39S}, \cite{2007A&A...465...51E}),
which was observed by MiNDSTEp and published by \cite{2013A&A...551A.104R},
and comparing the results for the two components with the results obtained in filter V and R by \cite{2012A&A...544A..51K}. The average weighted root mean square deviation
is $rms=1.6\sigma$.

\end{enumerate}

\subsection{Systematics with using GALFIT}

\begin{figure}[htbp]
\begin{center}
\includegraphics[scale=1,width=1.\columnwidth]{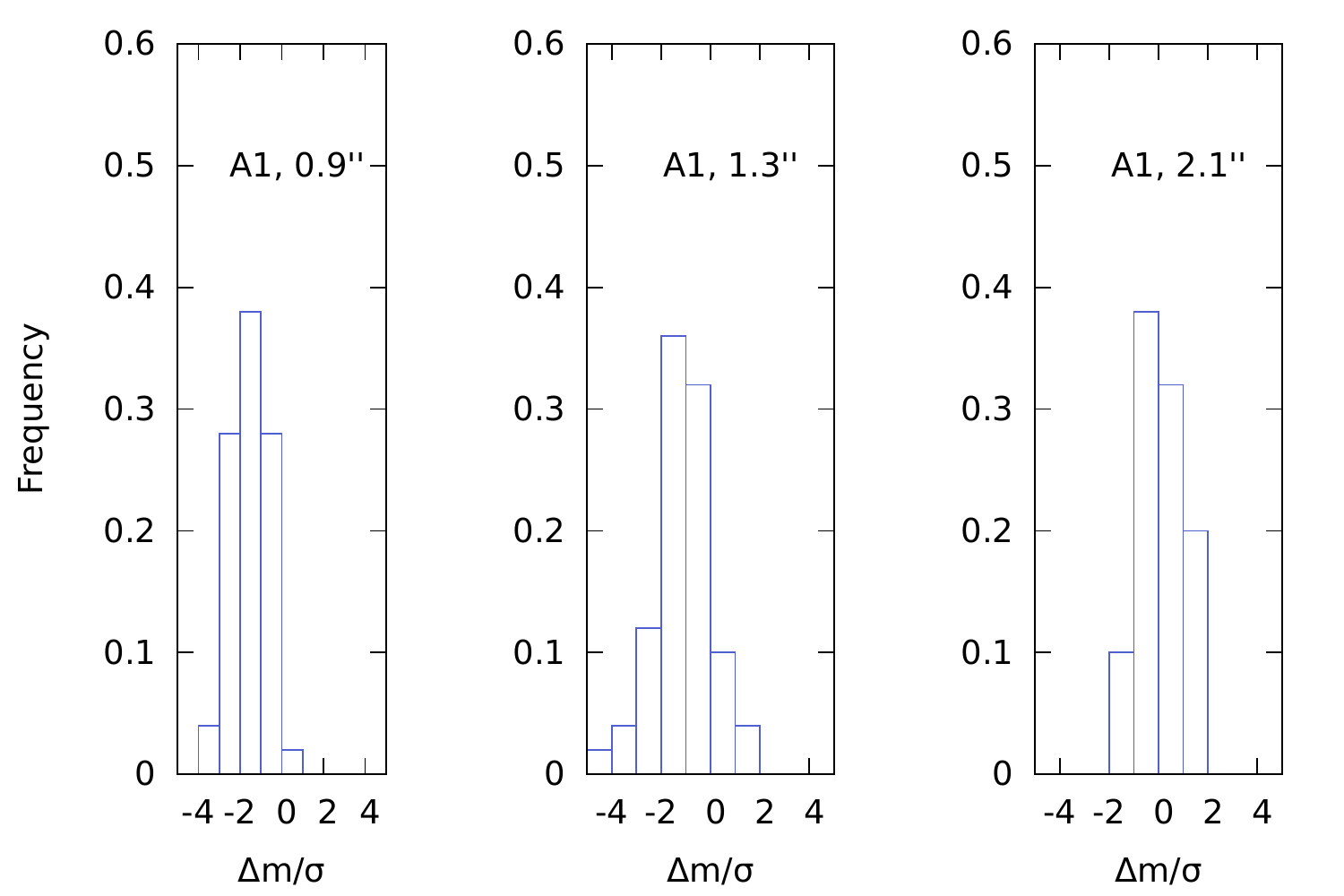}
\end{center}
\caption{Frequency distributions of $\Delta {m} / \sigma$; the difference between output and input magnitudes in units of GALFIT $\sigma$, for the component A1 of 50 mock models
of WFI\,2033-4723 under three different seeing regimes.}
\label{fig:errorbars_ass}
\end{figure}

In order to test the PSF fitting with GALFIT, we created mock models of the quasar 
WFI\,2033-4723 for three different values of the seeing in filter V. We chose to test the case of WFI\,2033-4723 because it is characterized by low fluxes and highly blended components.
Starting from three images of a real star in the surroundings of the quasar with FWHM
$0.9$, $1.31$, $2.1$ arcsec, using GALFIT we generated 50 artificial realizations of the quasar at each seeing value, with flux values
as computed from the V template and taking those into account as mean values of the corresponding Poissonian noise. We chose the quasar centroid at each realization within one pixel
with a uniform distribution. We added a sky background with mean value as computed from the V template to each artificial quasar,
including Poissonian noise, and a Gaussian readout noise realization. No lensing galaxy was included in these simulations.
The photometry of the artificial models was then carried out with GALFIT, choosing a PSF close to that used in building the artificial models.
Figure ~\ref{fig:errorbars_ass} shows the distribution of the ratio $\Delta {mag} / \sigma$,
which represents the difference between the GALFIT output flux and the known input flux in units of GALFIT sigmas (flux uncertainty) for quasar image A1 and for the three values
of seeing. The majority of realizations lies between $\Delta {mag}/\sigma\approx 0-2$ with minor tails at $\Delta {mag} / \sigma\approx 3$. We find similar
conclusions for the other images of the quasar. The effects, which led to the systematic discrepancy between
the input and output fluxes, are determined by the differences between the PSF of the quasar and the PSF chosen to model it. The average magnitude discrepancy
in the most frequent seeing regime does not exceed $0.02$ mag, which is negligible for the purposes of this paper.

\section{Results}
\label{sect:res}
\subsection{WFI\,2033-4723}
\begin{figure*}[htbp]
\begin{center}
\includegraphics[scale=1,width=1\textwidth]{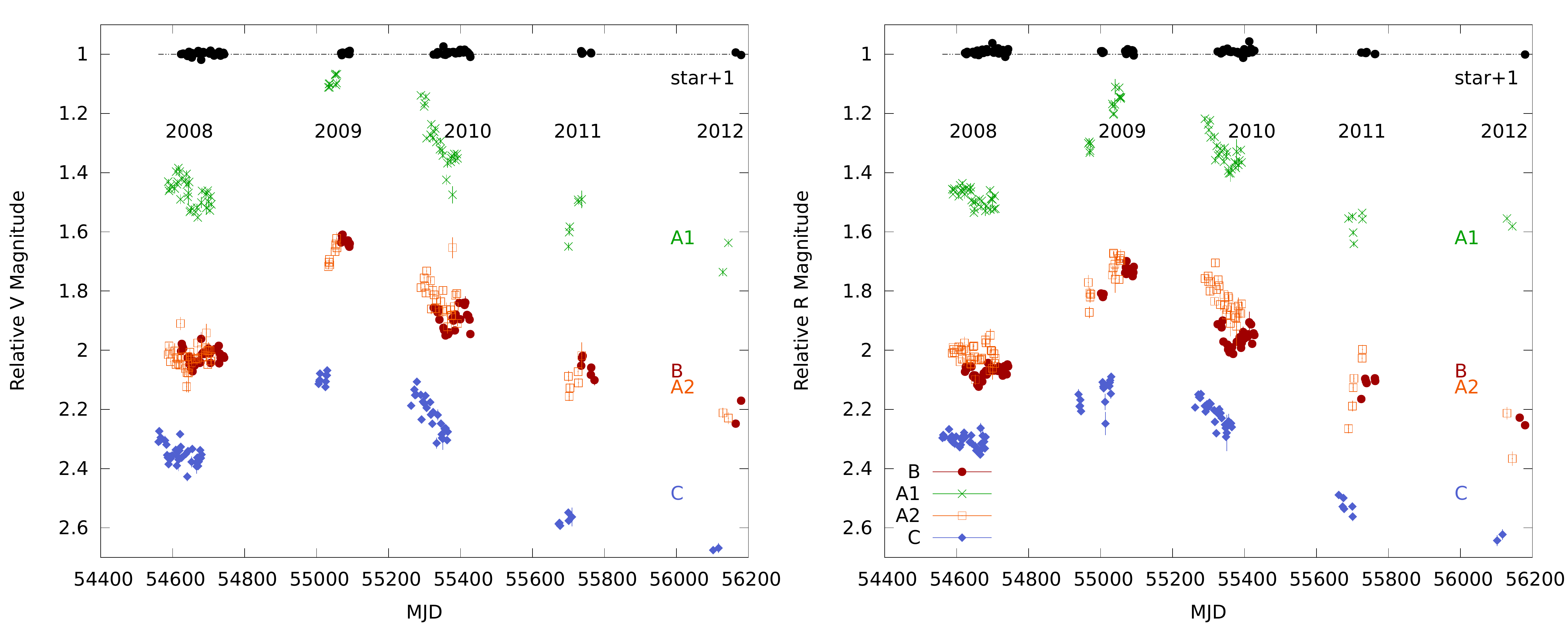}
\end{center}
\caption{V- (left) and R- (right) band light curves of WFI\,2033-4723 from 2008 to 2012. Components B (filled dots), A1 (asterisks), A2 (squares), C (filled diamonds)
are depicted in red, green, orange, and
blue, respectively. The light curve of a star in the field, labelled as Constant star/PSF in Fig.1, is shown in black and shifted down by 1 mag.} 
\label{fig:wfi2033V}
\end{figure*}

Figure ~\ref{fig:wfi2033V} shows the light curves for the quasar WFI\,2033-4723 components B, A1, A2, C,
and the constant star in Fig.~\ref{fig:wfi2033_stamps} in filters V and R,
respectively. The light curves are expressed by using
instrumental magnitudes, defined as
\begin{equation} m_X=-2.5\times\log_{10}(\frac{\Delta F_{X}+F_{X,T}}{F_{X,Ref}}),\end{equation}
where $\Delta F_{X}$ is the flux difference of the quasar components in the subtracted image, which has the photometric scale of the template, $F_{X,T}$ is the corresponding flux in the
template and $F_{X,Ref}$ that of the constant reference stars, indicated in Figs.~\ref{fig:wfi2033_stamps} and~\ref{fig:he0047_stamps}, where $X$ is V or R. Instrumental colours are defined accordingly.

The individual data points composing the light curves are also listed in Table~\ref{datawv}. The quoted error bars are determined by GALFIT, as explained
in paragraph 4.3.3. The light curves of images B (filled dots),
A1 (asterisks), A2 (squares), and C (filled diamonds) are shown in red, green, orange, and blue, respectively. The illustrated photometric data of the quasar components were shifted in accordance with
the time delays
measured by \cite{2008A&A...488..481V}, namely $\Delta t_{B-C}= 62.6$ days and $\Delta t_{B-A}=35.5$ days.

\begin{table}[htbp]
\large
  \caption{V-band and R-band yearly averages of WFI\,2033-4723 instrumental magnitudes.}
\label{tab:wfi2033yearly_mag}
\centering
\scalebox{0.7}{
\hskip-0.cm\begin{tabular}{p{1.cm}p{1.6cm}p{1.6cm}p{1.6cm}p{1.6cm}p{1.6cm}}
\hline
&2008&2009&2010&2011&2012\\
${(B)}_V$&$2.02\pm0.03$&$1.63\pm0.02$&$1.89\pm0.04$&$2.06\pm0.03$&$2.21\pm0.06$\\
${(A1)}_V$&$1.47\pm0.05$&$1.09\pm0.02$&$1.30\pm0.08$&$1.55\pm0.07$&$1.69\pm0.07$\\
${(A2)}_V$&$2.02\pm0.04$&$1.67\pm0.04$&$1.82\pm0.06$&$2.10\pm0.05$&$2.22\pm0.01$\\
${(C)}_V$&$2.35\pm0.03$&$2.10\pm0.02$&$2.22\pm0.06$&$2.58\pm0.02$&$2.67\pm0.01$\\
${(B)}_R$&$2.07\pm0.02$&$1.76\pm0.04$&$1.96\pm0.03$&$2.11\pm0.03$&$2.24\pm0.02$\\
${(A1)}_R$&$1.48\pm0.03$&$1.21\pm0.08$&$1.33\pm0.05$&$1.57\pm0.04$&$1.57\pm0.02$\\
${(A2)}_R$&$2.01\pm0.03$&$1.74\pm0.06$&$1.83\pm0.05$&$2.12\pm0.10$&$2.29\pm0.11$\\
${(C)}_R$&$2.30\pm0.02$&$2.15\pm0.04$&$2.22\pm0.04$&$2.52\pm0.03$&$2.63\pm0.02$\\
\hline
\end{tabular}}
\end{table}

\begin{figure*}[htbp]
\begin{center}
\includegraphics[scale=1,width=1\textwidth]{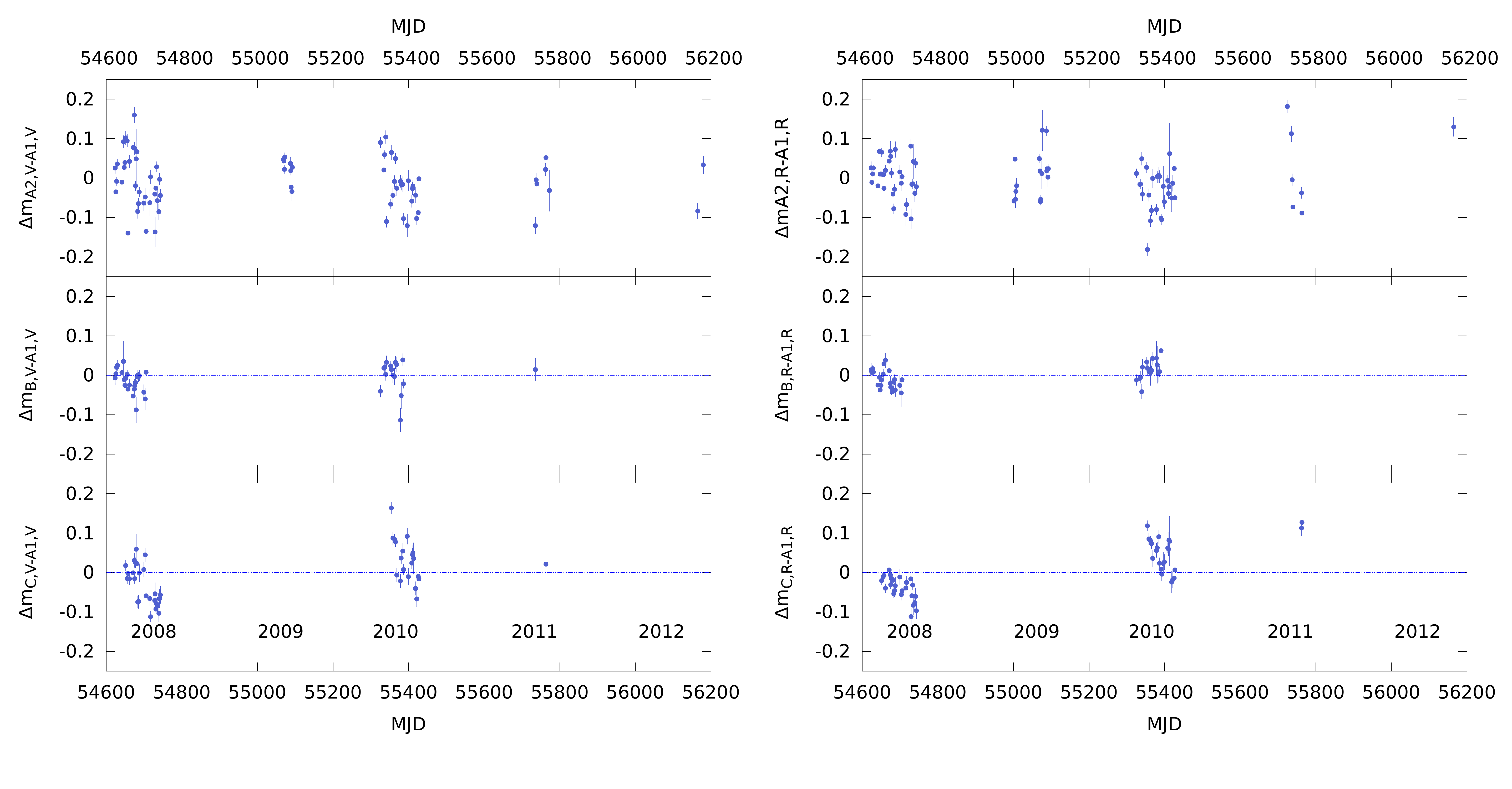}
\end{center}
\caption{Light curves A2-A1, B-A1, and C-A1 are shown in the V (left) and R band (right) in the upper, middle, and bottom panels, respectively.
The differences are computed, after correcting for the known time delays, by interpolating in between the data points of the brightest component
of each pair. The difference between C and A1 in the R band shows a significant magnitude variation of 0.16 mag between 2008 and 2011.}
\label{fig:wfi2033Val}
\end{figure*}

Components B, A1, A2, and C become brighter in 2009 and dimmer again across the remaining seasons, spanning a
magnitude range of $0.6$ mag in filter V and $\approx0.5$ mag in filter R. We list in Table~\ref{tab:wfi2033yearly_mag} 
the average magnitudes, computed within each season, for a more detailed picture of the brightness evolution of the four images. The quoted error
bars are standard deviations computed within each season.
Figure ~\ref{fig:wfi2033Val} is intended to reveal any differences between the variation of the four lensed components.
The light curves A2-A1, B-A1, and C-A1 are shown in the V and R bands. The differences were computed upon interpolation of the brightest component at
the observation dates of the weakest one in the pair, after correcting for the time delays between the pair components.
The figures suggest that the variation of component C through the observing seasons differs from the others, showing
a significant variation of $\sim0.16$ mag in the R band between 2008 and 2011.


\begin{figure*}[htbp]
\begin{center}
\includegraphics[scale=0.5,width=\textwidth,height=0.2\textheight]{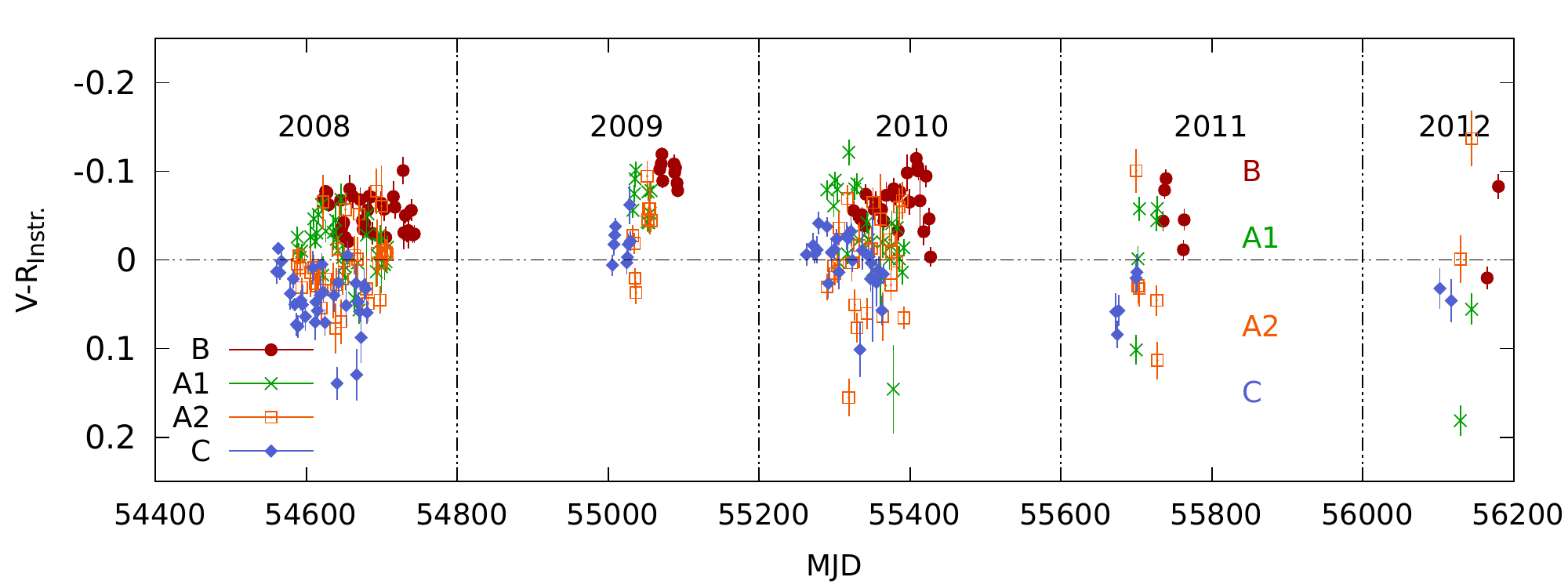}
\end{center}
\caption{$(V-R)_{Instr.}$ light curves of WFI\,2033-4723 from 2008 to 2012. Components B (filled dots), A1 (asterisks), A2 (squares), and C (filled diamonds) are depicted in red, green, orange, and blue, respectively.}
\label{fig:wfi2033color}
\end{figure*}

Figure ~\ref{fig:wfi2033color} shows the evolution of the colour $(V-R)_{Instr.}$ of the four components. The four images become bluer between 2008 and 2009 by $\approx 0.05$ mag in correspondence to the quasar brightening, and the images gradually become redder through the
remaining seasons.
We also compute the colour difference between the image pairs $A1-B$, $A2-B$, $C-B$, $A1-A2$, $A2-C$, and $A1-C$, after interpolating
the colour of the brightest component of each pair in correspondence to the days at which the other was observed.
In absence of lensing we expect the colours of the quasar components to differ only by
a constant, caused by the differential intergalactic extinction along their lines of sight and by the differential reddening by the lensing galaxy. When the colour light curves of the quasar images
cannot be matched by simply shifting them by a constant, the simplest explanation we can provide is microlensing
affecting images in an uncorrelated fashion (\cite{1992ApJ...397L...1W}).
We do not find systematic long-term colour difference variation across the whole observing campaign, but we cannot rule out intra-seasonal variations
of order $\approx0.1$mag. Intra-seasonal average values of the instrumental colour $(V-R)_{Instr.}$ for the four components and the colour difference between all possible
pairs of components are given in Tables ~\ref{tab:wfi2033yearlycolor} and ~\ref{tab:wfi2033yearlydiff}.

\begin{table}[htbp]
\large
  \caption{Yearly averages of the instrumental colour ${V-R}_{Instr.}$ for the four lensed components of quasar WFI\,2033-4723.}
\label{tab:wfi2033yearlycolor}
\centering
\scalebox{0.67}{
\hskip-0.3cm\begin{tabular}{p{1.9cm}p{1.9cm}p{1.9cm}p{1.9cm}p{1.9cm}p{1.9cm}}
\hline
&2008&2009&2010&2011&2012\\
${(V-R)_{Instr.}}_B$&$-0.05\pm0.02$&$-0.10\pm0.01$&$-0.06\pm0.03$&$-0.05\pm0.03$&$-0.03\pm0.07$\\
${(V-R)_{Instr.}}_{A1}$&$-0.01\pm0.03$&$-0.07\pm0.02$&$-0.03\pm0.05$&$-0.01\pm0.07$&$0.12\pm0.09$\\
${(V-R)_{Instr.}}_{A2}$&$0.01\pm0.04$&$-0.03\pm0.04$&$-0.01\pm0.08$&$0.02\pm0.08$&$-0.07\pm0.10$\\
${(V-R)_{Instr.}}_{C}$&$0.05\pm0.03$&$-0.02\pm0.02$&$0.00\pm0.03$&$0.05\pm0.03$&$0.04\pm0.01$\\
\hline
\end{tabular}}
\end{table}

\begin{table}[htbp]
\large
  \caption{Yearly average differences of the instrumental colour ${V-R}_{Instr.}$
for all the possible component pairs of quasar WFI\,2033-4723.}
\label{tab:wfi2033yearlydiff}
\centering
\scalebox{0.67}{
\hskip-0.4cm\begin{tabular}{p{2.3cm}p{1.9cm}p{1.9cm}p{1.9cm}p{1.7cm}p{1.7cm}}
\hline
&2008&2009&2010&2011&2012\\
$\overline{\Delta{(V-R)_{Instr.A1B}}}$&$0.05\pm0.03$& &$0.06\pm0.05$& & \\
$\overline{\Delta{(V-R)_{Instr.A2B}}}$&$0.05\pm0.04$& &$0.03\pm0.07$& &\\
$\overline{\Delta{(V-R)_{Instr.CB}}}$&$0.11\pm0.05$& &$0.08\pm0.03$& & \\
$\overline{\Delta{(V-R)_{Instr.A1A2}}}$&$-0.02\pm0.06$&$-0.04\pm0.06$&$0.0\pm0.1$&$0.0\pm0.1$&$0.19\pm0.01$\\
$\overline{\Delta{(V-R)_{Instr.A2C}}}$&$-0.05\pm0.04$& &$-0.01\pm0.05$& & \\
$\overline{\Delta{(V-R)_{Instr.A1C}}}$&$-0.06\pm0.04$& &$-0.04\pm0.04$& &\\
\hline
\end{tabular}}
\end{table}

\subsection{HE\,0047-1756}
\begin{figure*}[htbp]
\begin{center}
\includegraphics[scale=1,width=1\textwidth]{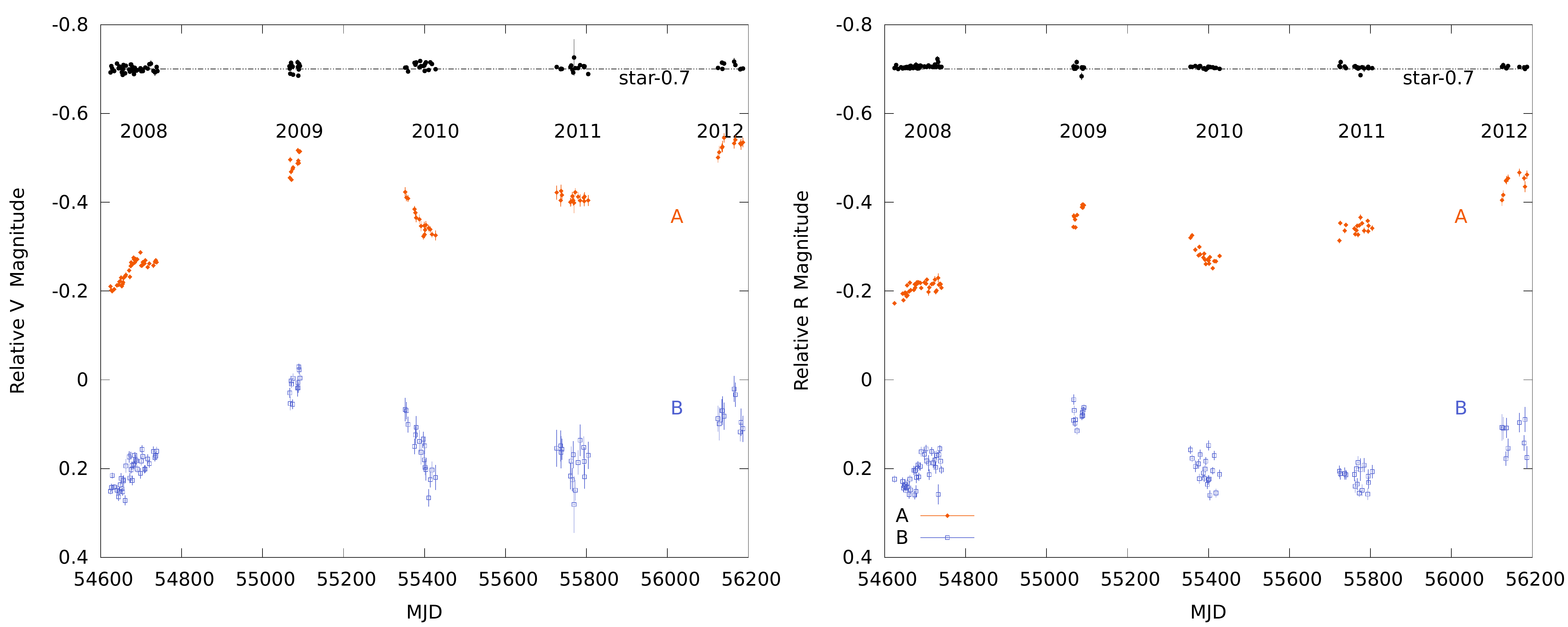}
\end{center}
\caption{V-(left) and R-(right) band light curves of HE\,0047-1756  from 2008 to 2012. Components A (filled dots) and B (squares) are depicted respectively in orange and blue.
The light curve of a star in the field, labelled as constant star in Fig.~\ref{fig:he0047_stamps}, is shown in black and shifted up by $-0.7$ mag.}
\label{fig:LChe0047V}
\end{figure*}

In Fig. ~\ref{fig:LChe0047V} we show the HE\,0047-1756 light curves during the years
2008-2012 for filters V and R. The individual data points composing the light curves are also listed in
Table~\ref{datahv}. The error bars are determined using GALFIT as explained in paragraph 3 of Sect. 4.3.

The light curves for images A and B are plotted in orange and blue, respectively, as a
function of the modified Julian date (MJD).
In 2008 the light curves of both lensed quasar images are characterized by a
 $\Delta m\approx 0.1$ mag intrinsic variation of the quasar
on timescales of $\approx50$ days. The data are consistent with
this rise ending around $MJD-2450000\approx
4680$ in image A, but around $MJD-2450000\approx 4690$ in image B. This delay of the brightness rise in image B is analysed in detail in Sect.~\ref{td}.
In the year 2009, both quasar images became brighter.
Starting from 2010 the quasar became dimmer and again brighter across
the last two periods. The overall amplitude of magnitude spanned by the light curves in both filters
does not exceed  $\approx 0.3$ mag. The average instrumental magnitudes of components A and B across the five periods
are shown in Table~\ref{tab:he0047yearly_mag}.

\begin{table}[htbp]
\large
  \caption{V-band and R-band yearly averages of the HE\,0047-1756 instrumental magnitudes.}
\label{tab:he0047yearly_mag}
\centering
\scalebox{0.67}{
\hskip0.cm\begin{tabular}{p{1.cm}p{1.9cm}p{1.9cm}p{1.9cm}p{1.9cm}p{1.9cm}}
\hline
&2008&2009&2010&2011&2012\\
${(A)}_V$&$-0.25\pm0.02$&$-0.49\pm0.02$&$-0.36\pm0.03$&$-0.41\pm0.01$&$-0.53\pm0.01$\\ 
${(B)}_V$&$1.21\pm0.03$&$1.01\pm0.03$&$1.16\pm0.06$&$1.19\pm0.04$&$1.08\pm0.03$\\
${(A)}_R$&$-0.21\pm0.01$&$-0.37\pm0.02$&$-0.28\pm0.02$&$-0.34\pm0.01$&$-0.44\pm0.02$\\
${(B)}_R$&$1.21\pm0.03$&$1.08\pm0.02$&$1.20\pm0.03$&$1.22\pm0.02$&$1.13\pm0.03$\\
\hline
\end{tabular}}
\end{table}

\subsection{Time delay for HE\,0047-1756}

\label{td}
\begin{figure}[htbp]
\begin{center}
\includegraphics[scale=1,width=\columnwidth]{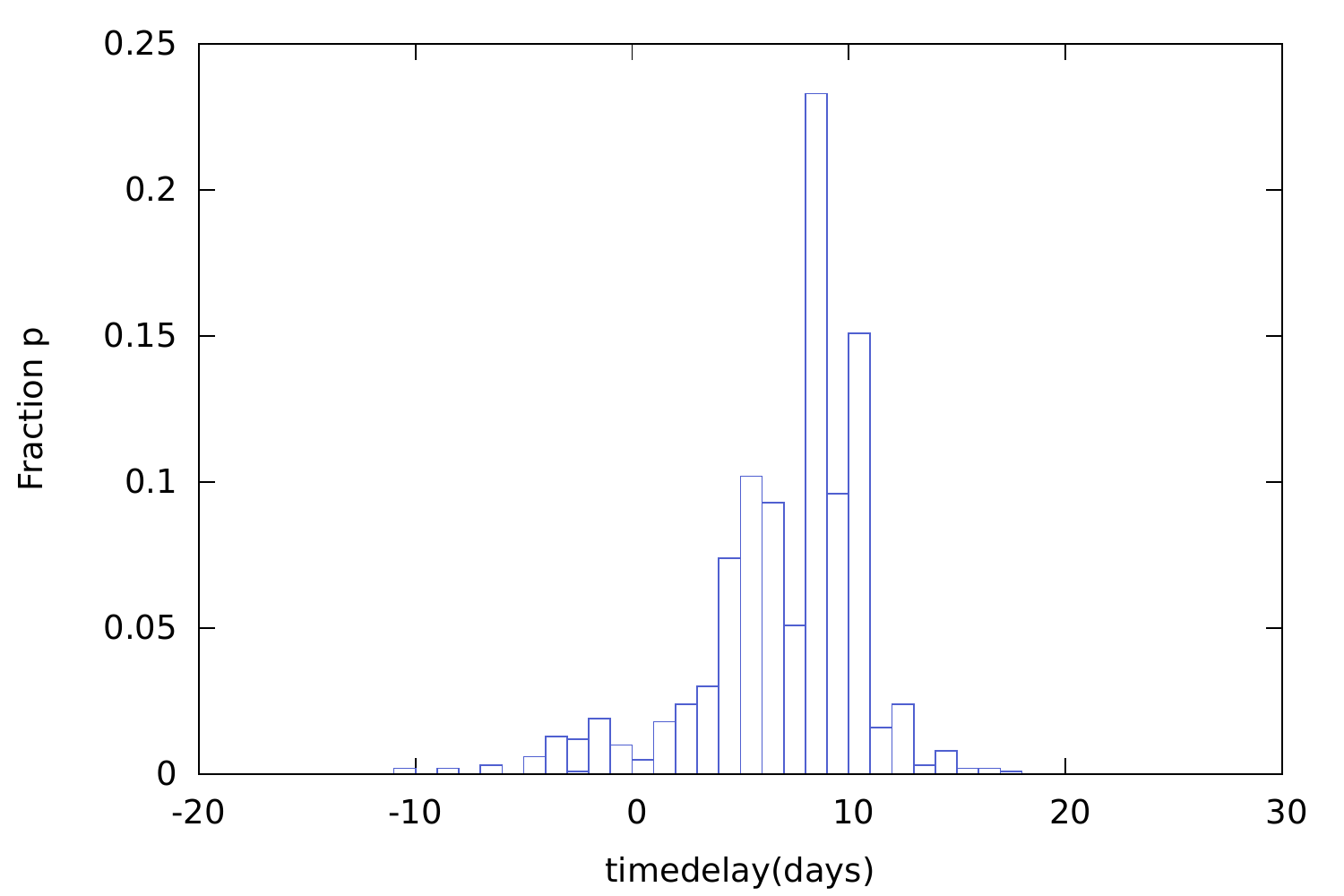}
\end{center}
\caption{Distribution $p$ for time delays days based on our light curves of components $A_{V}$ and $B_{V}$ obtained by applying the PyCS dispersion method.
The probability was computed from 1000 resamplings of the inferred intrinsic, extrinsic, and noise model.
Mean value and standard deviation of this distribution are $\Delta t= 8.0\pm4.2 $ days.}
\label{fig:he0047TDhisto_disp}
\end{figure}

\begin{figure}[htbp]
\begin{center}
\includegraphics[scale=1,width=\columnwidth]{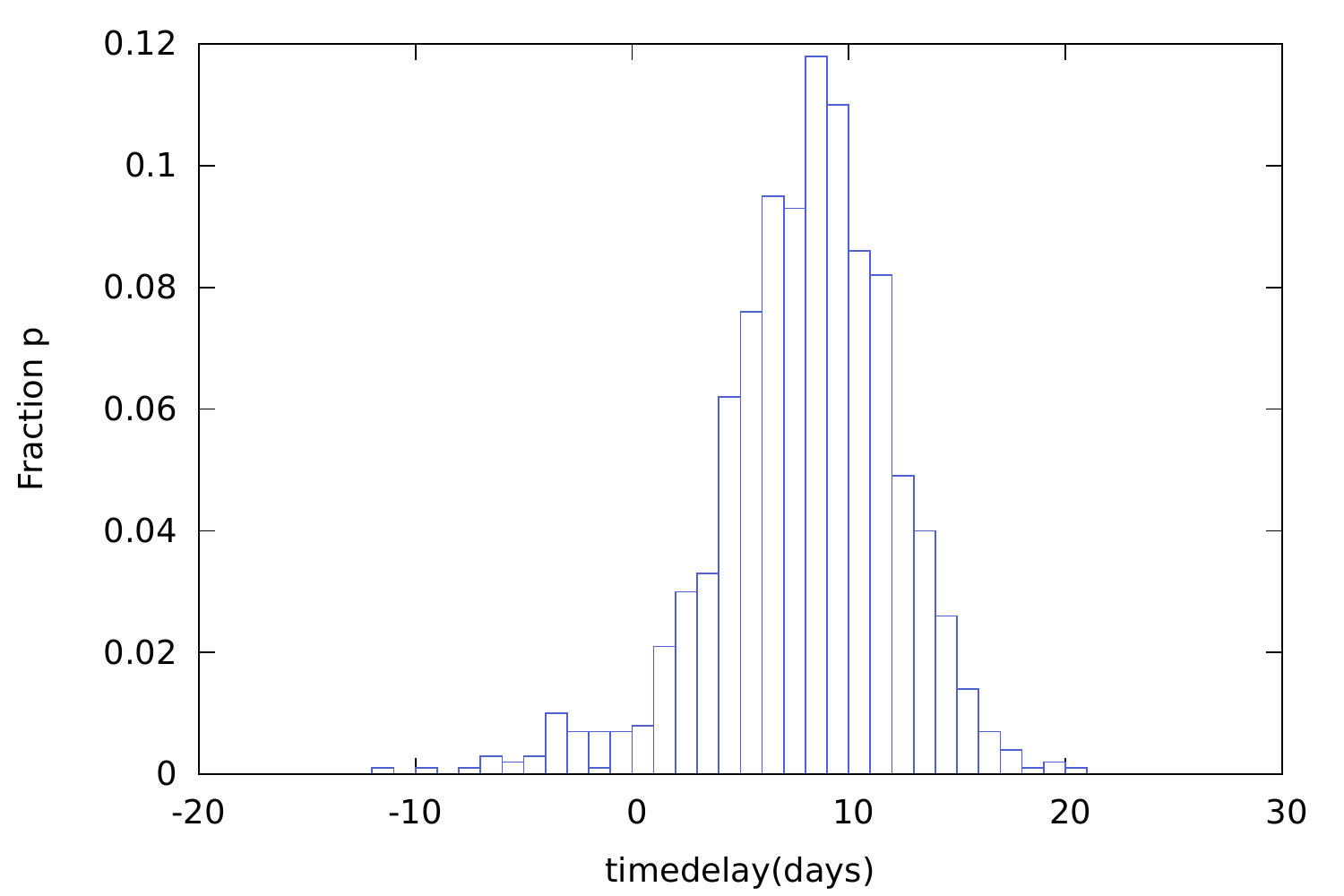}
\end{center}
\caption{Distribution $p$ for time delays days based on our light curves of components $A_{V}$ and $B_{V}$ obtained by applying the PyCS spline method.
The probability was computed from 1000 resamplings of the inferred intrinsic, extrinsic, and noise model.
Mean value and standard deviation of this distribution are $\Delta t= 7.2\pm3.8$ days.}
\label{fig:he0047TDhisto_spl}
\end{figure}

\begin{figure}[htbp]
\begin{center}
\includegraphics[scale=1,width=\columnwidth]{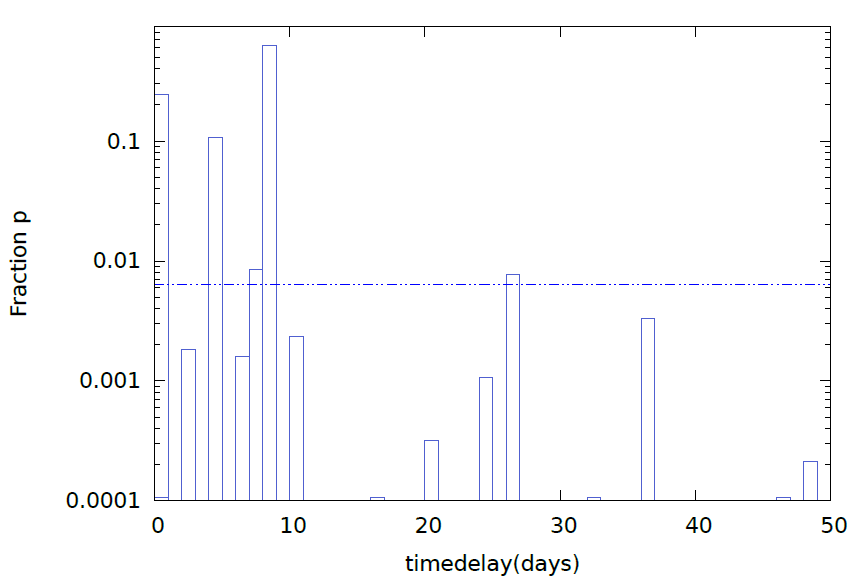}
\end{center}
\caption{Distribution $p$ for time delays between $0-50$ days based on our light curves of components $A_{V}$ and $B_{V}$.
The probability was computed from 10000 bootstrap resamplings of the observed light curves. For each resampling the brightest component A
was interpolated in correspondence to the dates at which B was observed. The region above the dashed line contains $95\%$ of the statistical weight of the distribution, after discarding the peak at 0 lag.
Mean value and standard deviation of this region are $\Delta t= 7.6\pm1.8$ days.}
\label{fig:he0047TDhisto}
\end{figure}

\begin{figure}[htbp]
\begin{center}
\includegraphics[scale=1,width=\columnwidth]{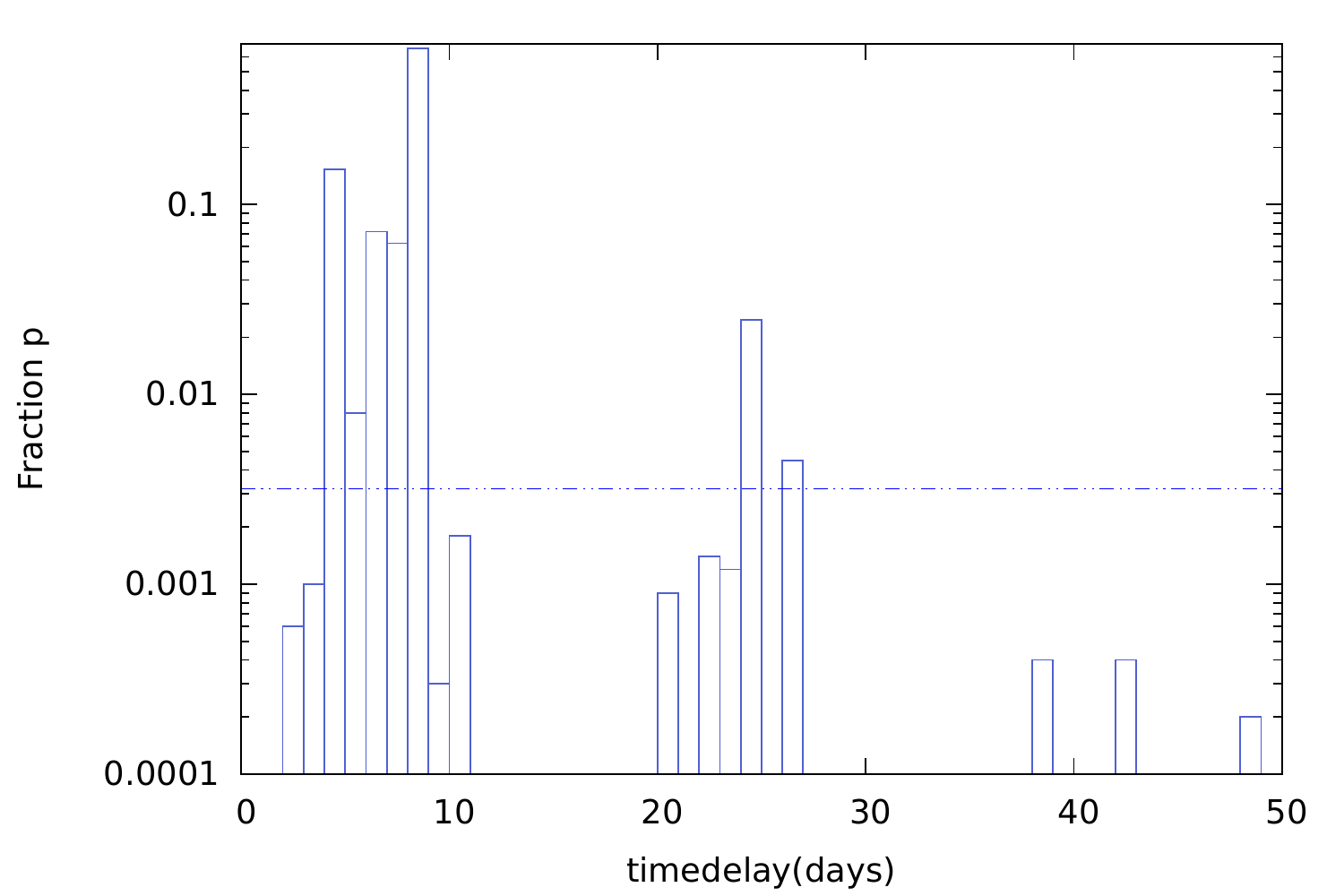}
\end{center}
\caption{Distribution $p$ for time delays between $0-50$ days based on our light curves of components $A_{V}$, $B_{V}$, and $B_{R}$.
The probability was computed from 10000 bootstrap resamplings of the observed light curves. For each resampling the brightest component A
was interpolated in correspondence to the dates at which B was observed. The region above the dashed line contains $95\%$ of the statistical weight.
Mean value and standard deviation of this region are $\Delta t= 7.6\pm2.9$ days.}
\label{fig:he0047TDhistomean}
\end{figure}

Several methods have been introduced to determine time delays in lensed systems
(\cite{1997ApJ...482...75K} and references therein; see also \cite{2001A&A...380..805B}, \cite{2002A&A...381..428G}, \cite{1996A&A...305...97P}, and \cite{2013A&A...553A.120T}).

\begin{figure}[htbp]
\begin{center}
\includegraphics[scale=1,width=\columnwidth]{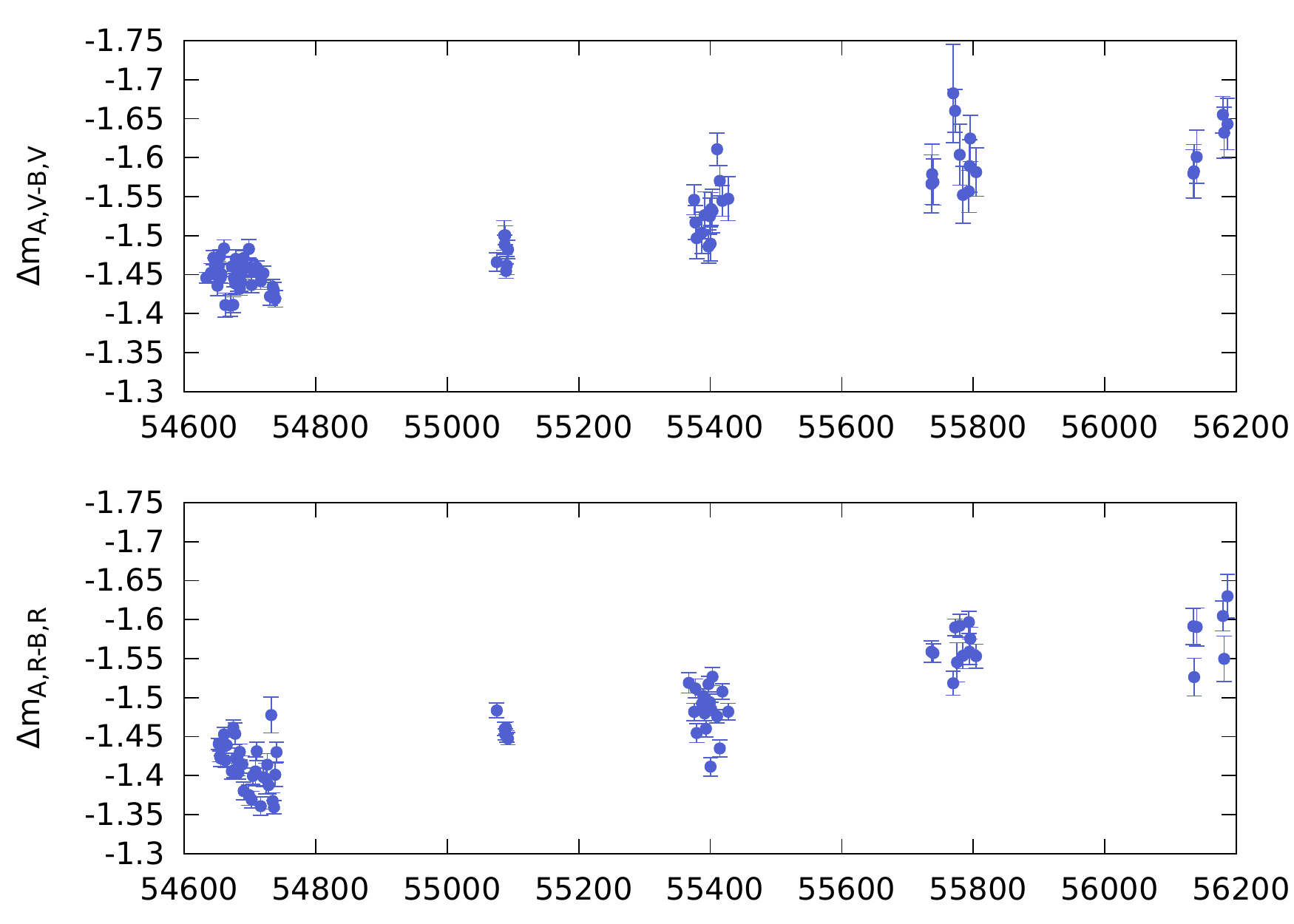}
\end{center}
\caption{Difference between the light curves of HE\,0047-1756 components A and B after shifting component A ahead by $7.6$ days. Component A is
interpolated at the epochs of component B before subtraction.}
\label{fig:he0047V_R_deltamag0812}
\end{figure}

Here, we apply the PyCS software by \cite{2013A&A...553A.120T} to our V and R
light curves from 2008 to 2012. This software allows for time delay measurements in presence of microlensing, defined as extrinsic variability, as opposed to the intrinsic variability of the quasar.
We use their free-knot spline technique and the dispersion technique. Appendix A summarizes the main input parameters for the PyCS spline and dispersion methods (see \cite{2013A&A...553A.120T}
for further details).\\
The first method uses splines to model both the intrinsic and extrinsic variability of the light curves and simultaneously adjusts the splines, time shifts, and magnitude shifts between the light curves to minimize a fitting
figure of merit involving all data points.\\
The second method goes back to the
dispersion techniques by \cite{1996A&A...305...97P} and simultaneously adjusts the time shifts
and low-order polynomial representations of the extrinsic variability to minimize a scalar dispersion function that quantifies the deviation between the light curves.
This method does not assume any model for the intrinsic variability.\\
The results determined with the spline fitting technique and dispersion technique in the V band are $7.2\pm3.8$ days and $8.0\pm4.2$ days, respectively, with image A leading.
The mean value and quoted error bars
correspond to the mean and standard deviation of the
resulting time delay distributions, which are obtained by drawing 1000 realizations of the observed light curves; these are shown in Figs. ~\ref{fig:he0047TDhisto_disp} and  ~\ref{fig:he0047TDhisto_spl}. 
The light curve realizations are drawn taking into account a model for the intrinsic variability, a model for extrinsic variability, and a noise model, as explained in \cite{2013A&A...553A.120T}.
On the other hand,
the application of these techniques to the R-band data in the years 2008-2012 does not converge on a unique answer.
In the following we carry out a zoom-in analysis on the year 2008 in the V band only to confirm the obtained results and show that the above analysis was not biased by the existence of the observing gaps. We found out that the best way to analyse such a short light curve portion is to follow a linear interpolation scheme, which does not introduce the difficulties of generating the model for the intrinsic and extrinsic variability, which are necessary inputs for the PyCS  software to draw new realizations of the light curves, from a smaller number of data points.
Our approch here, aimed at determining the time delay that minimizes
the magnitude difference between the light curves, was published first by \cite{2005A&A...440...53G} and goes back to
\cite{1997ApJ...482...75K}. It consists of the following steps:

\begin{enumerate}
\item  Component A of each of 10000 bootstrap resamplings of the observed light curve is shifted by the time delay 
$\Delta t$ to be tested, whose values lie in the range from 0 to 50 days (image A leading). Such a high number of resamplings constrains the uncertainties
on the time delay measurement. 
\item The light curve of the brightest component A is linearly interpolated to match the dates at which the B light curve
has been observed. Only gaps shorter than 20 days are interpolated. 
\item Each resampling is smoothed by triangular filter with a width of 3 and 6 days for the brightest component A and the weakest component B, respectively.
The use of larger windows, i.e. 6 and 12 days, does not change the results.
\item From the resulting light curves, comprising N epochs, we compute the weighted magnitude difference between the components $\Delta m$ and the $\chi^2_{\nu}$

\begin{equation}\chi^2_{\nu} = \frac{1}{N-2}\sum_{1}^{N} \frac{(m_A(t_i)-m_B(t_i+\Delta t)-\Delta m)^2}{{\sigma_{mA}}^2+{\sigma_{mB}}^2}  \end{equation}

is determined, where we call $t_i$ the generic time at which data has been collected. The parameters ${\sigma_{mA}}$ and ${\sigma_{mB}}$ are the 
Poissonian noise propagated through the interpolation formula for $m_A(t_i)$ and the Poissonian noise corresponding to $m_B(t_i+\Delta t)$, respectively.
\item The time delay corresponding to the minimum $\chi^2_{\nu}$ is the optimal time delay at any given resampling. 
\end{enumerate}
The algorithm is applied to the light curve couple $A_VB_{V}$.
The probability distribution of time delays obtained using this method is plotted in Fig. ~\ref{fig:he0047TDhisto}. The probability of each $1$-day bin is calculated as the ratio
between the occurence of light curves with best-fitting time delay in that bin and the total number of resamplings.
This procedure also produces a peak for 0 lag, which might be interpreted as a false peak, that is derived from correlated brightness fluctuations at 0 lag when dealing with optical discrete data, also reported by \cite{2006A&A...447..905V} and  \cite{2003ApJ...587...71C}, who describes it as a ``frame-to-frame correlation
error in the photometry''.
We compute mean time delay and standard deviation for the region above the dashed line, which carries $95\%$ of the statistical weight of the distribution (not taking into account the 0 lag peak), 
and obtain $\Delta t= 7.6\pm1.8$ days. In order to assess whether the distribution at 0 lag corresponds to a false peak, we apply the above algorithm to the 2008 light curve couple $A_VB_{VR}$, where $B_{VR}$ is the average of the light curves of component B in both filters. The aim is to break the 0 lag correlation between the multiple photometric data recorded on a single frame. This is not strictly correct because interband time delays have been measured for several non-lensed quasars (\cite{2010MmSAI..81..138K}) as due to light travel time differences between two different emission regions and, in addition, the above full light curve analysis in the R band has not converged to a unique value. However, the procedure leads to a time delay distribution (shown in Fig. ~\ref{fig:he0047TDhistomean}) with no peak at 0 lag, whose $95\%$ statistical weight region is described by a mean time delay and standard deviation of $\Delta t= 7.6\pm2.9$ days. Therefore, we conclude that the time delay analysis carried out by only taking the year 2008 into account in the V band produces a result that is consistent with the above analysis including the full light curves. 
In Fig.~\ref{fig:he0047V_R_deltamag0812} we compute the difference between the light curves of components A and B in both filters. This is carried out after shifting component A ahead by $7.6$ days and
interpolating it at the epochs of component B. From these plots a secular evolution of the magnitude difference between the quasar components can be seen. This evolution of $\approx 0.2$ mag across the five periods is mainly linear; we explain it as due to a long-term microlensing perturbation.
Such a behaviour has already been observed for the double quasar SBS 1520+530 by \cite{2005A&A...440...53G}.

\begin{figure}[htbp]
\begin{center}
\includegraphics[scale=1,width=\columnwidth]{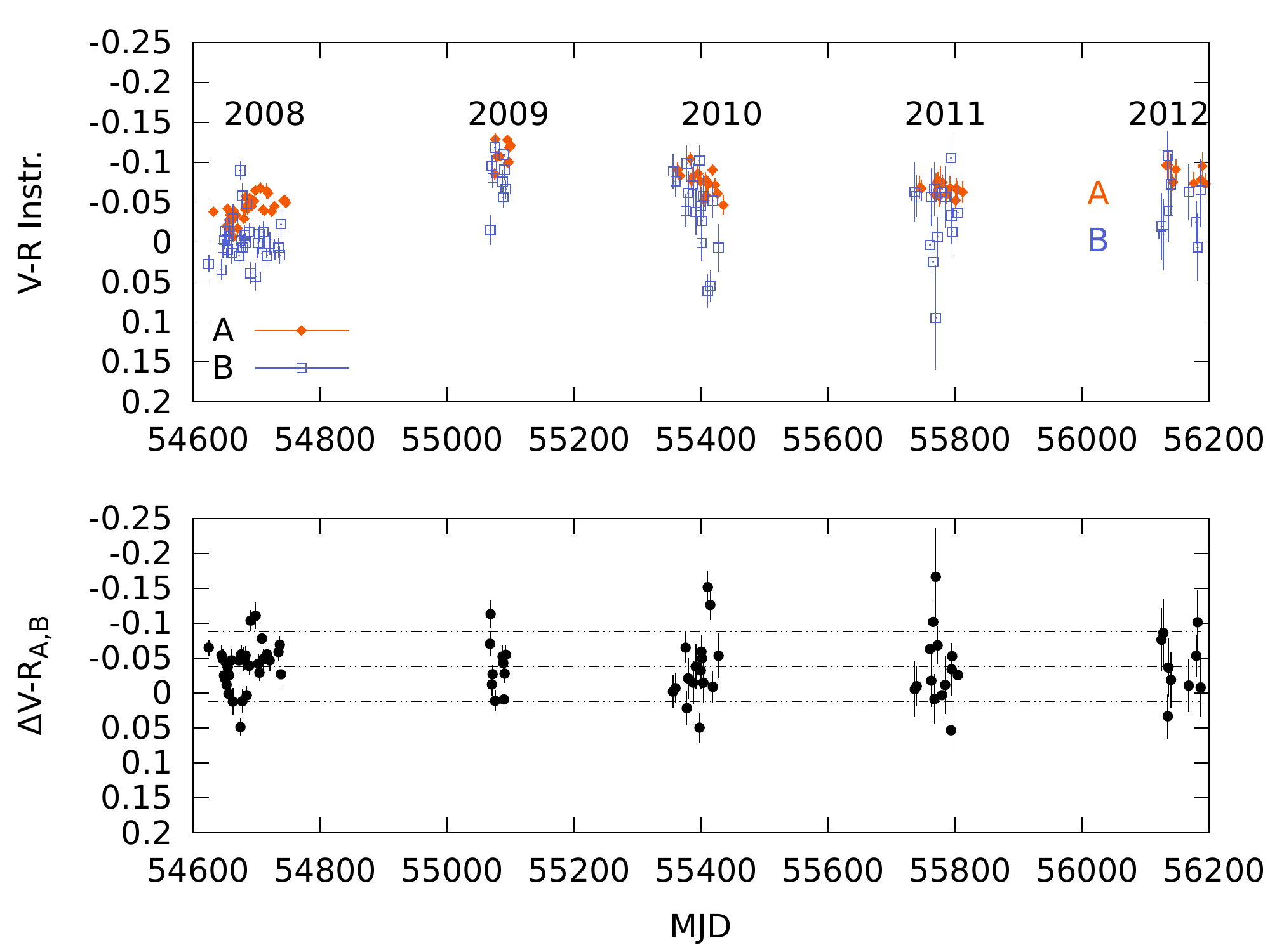}
\end{center}
\caption{$(V-R)_{Instr.}$ light curves of HE\,0047-1756 from 2008 to 2012. Components A (diamonds) and B (squares) are depicted in orange and blue, respectively, in the first upper panel.
In the bottom panel we show how the colour difference between the two components evolves. The difference 
is computed by interpolating the magnitude of the brightest component in the pair in correspondence to the dates at which the weakest one has been observed.
The horizontal dashed lines define $\pm0.05$ intervals around the average colour difference.}
\label{fig:he0047color}
\end{figure}

An analysis of the colour index ${V-R}_{Instr.}$ light curve as a function of the MJD, shown in the upper part of
Fig.~\ref{fig:he0047color}, reveals that the two components span the highest
colour variation of $\approx 0.07$ mag between seasons 2008 and 2009, with both images turning bluer in
correspondence to the 2009 brightening, as already observed by \cite{2004ApJ...601..692V}, \cite{2006ApJ...642...87P}, \cite{2011A&A...528A..42R}, and \cite{2013A&A...551A.104R}. The yearly averages of the colour are shown in Table~\ref{tab:he0047yearlycolor}.
\begin{table}[htbp]
\large
  \caption{Yearly averages of the instrumental colour ${V-R}_{Instr.}$ for the two lensed components of quasar HE\,0047-1756.}
\label{tab:he0047yearlycolor}
\centering
\scalebox{0.67}{
\hskip-0.4cm\begin{tabular}{p{1.9cm}p{1.9cm}p{1.9cm}p{1.9cm}p{1.9cm}p{1.9cm}}
\hline
&2008&2009&2010&2011&2012\\
${(V-R)_{Instr.}}_A$&$-0.04\pm0.02$&$-0.11\pm0.01$&$-0.08\pm0.02$&$-0.07\pm0.01$&$-0.08\pm0.01$\\
${(V-R)_{Instr.}}_B$&$0.00\pm0.03$&$-0.07\pm0.04$&$-0.04\pm0.05$&$-0.03\pm0.05$&$-0.04\pm0.04$\\
\hline
\end{tabular}}
\end{table}

\begin{table}[htbp]
\large
  \caption{Yearly average differences of the instrumental colour ${V-R}_{Instr.}$ between components A and B of quasar HE\,0047-1756.}
\label{tab:he0047yearlydiff}
\centering
\scalebox{0.67}{
\hskip-0.4cm\begin{tabular}{p{1.9cm}p{1.9cm}p{1.9cm}p{1.9cm}p{1.9cm}p{1.9cm}}
\hline
&2008&2009&2010&2011&2012\\
$\overline{\Delta{(V-R)_{Instr.}}}$&$-0.04\pm0.04$&$-0.03\pm0.03$&$-0.05\pm0.05$&$-0.04\pm0.06$&$-0.03\pm0.04$\\
\hline
\end{tabular}}
\end{table}

The colour difference between the quasar images is mainly constant $\approx 0.04$ throughout the 
five seasons, as shown in the bottom panel of Fig.~\ref{fig:he0047color}. The colour difference
was computed after shifting the colour light curve of component A by  $7.6$ days and linearly interpolating it
in correspondence to the observation dates of component B.
The yearly averages of the colour difference are shown in Table~\ref{tab:he0047yearlydiff} and are consistent
with a constant offset between the colour light curves.

\section{Summary and discussion}
\label{sect:summ}

We have presented $V$-band and $R$-band photometry of the gravitational
lens systems WFI\,2033-4723 and HE\,0047-1756 from 2008 to 2012,
based on data collected by MiNDSTEp with the Danish 1.54 m at the ESO La Silla observatory, Chile.
By applying the Alard \& Lupton image subtraction method (\cite{1998ApJ...503..325A}, \cite{2000A&AS..144..363A}) we have constructed
the light curves of the quasar components.

\begin{enumerate}
\item The lensed images of WFI\,2033-4723 vary by $0.6$ mag in V and $\approx0.5$ mag in R
during the campaign, becoming brighter in 2009 and gradually weaker until 2012. 
After computing the A2-A1, B-A1, C-A1 light curves, we note that C-A1 shows a variation of $\approx 0.2$mag in the R band across seasons 2008-2011.
We suggest that microlensing that only affects the outer part of the accretion disk of image C could in
principle explain the behaviour seen in the R band; this relies on the hypothesis that an outer and hence cooler part of the disk, with emission at longer
wavelengths, is magnified by the caustic pattern.\\

\item The two lensed components of quasar HE\,0047-1756 reach their maximum brightness in 2009 and again in 2012 with a magnitude variation of $\approx 0.2-0.3$ mag,
depending on which components and filters are considered.
For the first time we provide a measurement of the time delay between the two components. We apply the PyCS software by \cite{2013A&A...553A.120T} to our whole V-band and R-band data set.
The free-knot spline technique and dispersion technique provide consistent estimates of the time delay of $7.2\pm3.8$ and $8.0\pm4.2$ days in the V band. 
On the other hand, the two techniques do not converge to a unique result in the R band.
By making use of a linear-interpolation scheme applied to 
the brightest component A (see \cite{2005A&A...440...53G}), we carry out a zoom-in analysis on the year 2008 in the V band and find that the time delay value minimizing the magnitude difference between the light curves $A_V$ and $B_{V}$ in 2008 is $\Delta t= 7.6\pm1.8$ days, which is consistent with the above results.
The magnitude difference between the light curves of A and B in both bands increases from 2008 to 2012 by $\approx0.2$ mag, showing a long-term linear uncorrelation
between the two components, which can be explained with a long-term microlensing perturbation.\\

The images of both quasars become bluer
when getting brighter. This is consistent with previous studies (e.g. \cite{2004ApJ...601..692V}).
A simple possible explanation to this is obtained by considering the accretion disk models for quasars. A boost
in the disk accretion rate
could produce a temperature increase of the inner regions of a quasar, hence a brighter and bluer emission.
The colour difference between the components of each quasar
is consistent with being constant across the five periods.
\end{enumerate}
\begin{acknowledgements}
We would like to thank the anonymous referee for having significantly contributed to improving the quality of this manuscript.
We would like to thank Armin Rest for introducing us to the HoTPANnTS software.
We also thank Ekaterina Koptelova for having provided the light curves of quasar UM673.
EG gratefully acknowledges the support of the International Max Planck Research School for Astrophysics (IMPRS-HD)
and the HGSFP. EG also thanks Katie Ramir\'e for helpful suggestions.
TA acknowledges support from FONDECYT proyecto 11130630 and the Ministry of Economy, Development, and Tourism's Millennium Science Initiative through grant IC120009, awarded to The Millennium Institute of Astrophysics, MAS.
MD and MH are supported by NPRP grant NPRP-09-476-1-78 from the Qatar National Research
Fund (a member of Qatar Foundation). MH acknowledges support from the Villum foundation.
This publication was made possible by NPRP grant \# X-019-1-006 from the Qatar National Research Fund (a member of Qatar Foundation).
The research leading to these results has received funding from the European Union Seventh Framework Programme (FP7/2007-2013) under grant agreement no. 268421.
TCH would like to acknowledge financial support from KASI travel grant 2012-1-410-02 and Korea Research Council for Fundamental Science and Technology (KRCF).
DR acknowledges financial support from the Spanish Ministry of
Economy and Competitiveness (MINECO) under the 2011 Severo Ochoa
Program MINECO SEV-2011-0187.
Funding for the Stellar Astrophysics Centre is provided by The Danish National Research Foundation (Grant agreement no.: DNRF106). The research is supported by the ASTERISK project (ASTERoseismic Investigations with SONG and Kepler) funded by the European Research Council (Grant agreement no.: 267864).
YD, AE, FF, DR, OW, and J. Surdej acknowledge support from the Communaut\'e fran\c{c}aise de Belgique - Actions de recherche 
concert\'ees - Acad\'emie Wallonie-Europe.
\end{acknowledgements} 

\clearpage
\bibliographystyle{aa} 
\bibliography{biblio}

\onecolumn
\fontsize{7}{9}\selectfont{\begin{longtable}{ccccccccccccccccc}

\caption{\label{datawv} V- and R- band photometry of  WFI\,2033-4723 , as in Fig.~\ref{fig:wfi2033V}.}\\

\hline
\hline

mag $B_V$ & $\sigma_{B,V}$ & mag $A1_V$ & ${\sigma}_{A1,V}$ & mag $A2_V$ & ${\sigma}_{A2,V}$ & mag $C_V$ & $\sigma_{C,V}$ &mag $B_R$ & $\sigma_{B,R}$ & mag $A1_R$ & ${\sigma}_{A1,R}$ & mag $A2_R$ & ${\sigma}_{A2,R}$ & mag $C_R$ & $\sigma_{C,R}$ &$MJD$\\
\hline 
\endfirsthead 
\caption{continued.}\\ 
 
\hline
\hline
mag $B_V$ & $\sigma_{B,V}$ & mag $A1_V$ & ${\sigma}_{A1,V}$ & mag $A2_V$ & ${\sigma}_{A2,V}$ & mag $C_V$ & $\sigma_{C,V}$ &mag $B_R$ & $\sigma_{B,R}$ & mag $A1_R$ & ${\sigma}_{A1,R}$ & mag $A2_R$ & ${\sigma}_{A2,R}$ & mag $C_R$ & $\sigma_{C,R}$ &$MJD$\\
\hline 
\endhead 
\hline
\endfoot
2.003	&0.006	&1.43	&0.007	&2.015	&0.011	&2.31	&0.009 &2.073	        &0.007	&1.456	&0.008	&2.01	&0.013	&2.296	&0.009	&54623.4\\
1.979	&0.004	&1.461	&0.006	&1.985	&0.008	&2.273	&0.005	 & 2.056	&0.002	&1.473	&0.002	&1.99	&0.002	&2.286	&0.002	&54625.4\\
1.988	&0.004	&1.456	&0.006	&2.006	&0.009	&2.306	&0.007	 & 2.064	&0.002	&1.459	&0.002	&1.997	&0.003	&2.291	&0.002	&54627.4\\
1.994	&0.006	&1.444	&0.007	&2.039	&0.01	&2.295	&0.008	 & 2.056	&0.004	&1.454	&0.003	&2.008	&0.005	&2.293	&0.004	&54629.4\\
2.026	&0.01	&1.454	&0.016	&2.003	&0.025	&2.305	&0.015	 & 2.056	&0.007	&1.48	&0.008	&1.988	&0.012	&2.267	&0.01	&54641.4\\
2.021	&0.008	&1.397	&0.009	&2.048	&0.013	&2.319	&0.009	 & 2.088	&0.004	&1.443	&0.004	&2.039	&0.005	&2.298	&0.004	&54645.4\\
2.047	&0.005	&1.439	&0.006	&2.025	&0.008	&2.354	&0.006	 & 2.085	&0.004	&1.46	&0.003	&1.998	&0.005	&2.304	&0.003	&54647.4\\
2.051	&0.007	&1.431	&0.008	&2.029	&0.013	&2.363	&0.009	 & 2.093	&0.006	&1.461	&0.007	&2	&0.012	&2.29	&0.008	&54649.4\\
2.06	&0.007	&1.385	&0.009	&2.046	&0.014	&2.385	&0.01	 &2.085	&0.005	&1.436	&0.006	&2.03	&0.009	&2.31	&0.006	&54651.4\\
2.021	&0.017	&1.473	&0.03	&1.99	&0.046	&2.387	&0.029	 & & & & & & & & &54653.4\\
2.071	&0.008	&1.396	&0.009	&2.05	&0.013	&2.359	&0.01	 & 2.092	&0.004	&1.459	&0.003	&1.996	&0.004	&2.313	&0.004	&54655.4\\
2.037	&0.011	&1.49	&0.015	&1.91	&0.022	&2.366	&0.016	 & 2.117	&0.011	&1.473	&0.014	&1.976	&0.021	&2.316	&0.015	&54657.4\\
2.051	&0.009	&1.421	&0.009	&2.022	&0.013	&2.356	&0.012	 & 2.123	&0.009	&1.453	&0.008	&2.001	&0.011	&2.292	&0.01	&54661.3\\
2.038	&0.009	&1.426	&0.013	&2.062	&0.022	&2.337	&0.013	 & 2.106	&0.009	&1.457	&0.014	&2.028	&0.022	&2.328	&0.013	&54671.3\\
2.04	&0.009	&1.404	&0.01	&2.123	&0.018	&2.39	&0.012	 & 2.084	&0.011	&1.449	&0.013	&2.046	&0.021	&2.319	&0.015	&54674.3\\
2.043	&0.007	&1.441	&0.009	&2.075	&0.013	&2.344	&0.009	 & 2.078	&0.007	&1.464	&0.008	&2.047	&0.012	&2.296	&0.008	&54675.3\\
2.041	&0.006	&1.482	&0.007	&2.021	&0.01	&2.36	&0.008	 & 2.083	&0.006	&1.493	&0.007	&2.034	&0.009	&2.303	&0.006	&54677.3\\
1.962	&0.022	&1.469	&0.038	&2.076	&0.066	&2.37	&0.038	& & & & & & & & &54679.3\\
2.014	&0.014	&1.432	&0.015	&2.057	&0.022	&2.339	&0.019	 &  2.07	&0.006	&1.5	&0.007	&1.988	&0.009	&2.3	&0.007	&54681.3\\
2.011	&0.008	&1.533	&0.01	&2.007	&0.014	&2.284	&0.009	 &  2.082	&0.006	&1.535	&0.008	&1.986	&0.01	&2.279	&0.008	&54683.3\\
2.008	&0.006	&1.529	&0.008	&2.023	&0.011	&2.327	&0.007	 &  2.079	&0.006	&1.523	&0.007	&2.022	&0.01	&2.29	&0.008	&54685.3\\
2.013	&0.004	&1.519	&0.007	&2.043	&0.01	&2.364	&0.008	 &  2.043	&0.011	&1.499	&0.011	&2.099	&0.017	&2.293	&0.013	&54687.3\\
1.995	&0.009	&1.533	&0.011	&2.028	&0.015	&2.35	&0.013	 &  2.059	&0.009	&1.49	&0.01	&2.033	&0.015	&2.31	&0.012	&54699.3\\
2.012	&0.01	&1.519	&0.013	&2.03	&0.019	&2.427	&0.014	 &  2.069	&0.008	&1.515	&0.01	&2.031	&0.016	&2.288	&0.011	&54703.3\\
2.043	&0.008	&1.552	&0.011	&1.975	&0.015	&2.341	&0.011	 &  2.069	&0.008	&1.495	&0.01	&2.027	&0.015	&2.315	&0.011	&54705.3\\
1.999	&0.011	&1.501	&0.018	&1.998	&0.028	&2.378	&0.018	 &  2.07	&0.012	&1.529	&0.016	&1.965	&0.023	&2.326	&0.016	&54715.2\\
1.997	&0.008	&1.462	&0.01	&2.024	&0.014	&2.334	&0.011	 &  2.056	&0.009	&1.514	&0.01	&1.975	&0.013	&2.339	&0.011	&54717.2\\
1.985	&0.012	&1.473	&0.013	&1.992	&0.02	&2.379	&0.017	 &  2.086	&0.009	&1.46	&0.01	&2.069	&0.016	&2.353	&0.013	&54728.2\\
2.044	&0.014	&1.519	&0.022	&1.942	&0.03	&2.393	&0.024	 &  2.075	&0.011	&1.525	&0.015	&1.949	&0.021	&2.263	&0.016	&54729.2\\
2.012	&0.006	&1.47	&0.006	&2.003	&0.008	&2.363	&0.008	 &  2.062	&0.005	&1.488	&0.006	&2	&0.008	&2.316	&0.006	&54731.2\\
2.025	&0.005	&1.461	&0.007	&2.048	&0.011	&2.391	&0.008	 &  2.054	&0.005	&1.489	&0.007	&2.003	&0.01	&2.333	&0.007	&54733.2\\
2.024	&0.012	&1.498	&0.017	&2	&0.026	&2.376	&0.018	 &  2.057	&0.015	&1.491	&0.023	&2.061	&0.038	&2.289	&0.022	&54735.1\\
2.025	&0.009	&1.528	&0.011	&2.002	&0.016	&2.338	&0.011	 &  2.081	&0.008	&1.523	&0.012	&2.012	&0.017	&2.309	&0.014	&54739.1\\
2.02	&0.006	&1.481	&0.008	&2.036	&0.012	&2.365	&0.008	 &  2.048	&0.006	&1.479	&0.007	&2.045	&0.009	&2.332	&0.007	&54741.1\\
2.025	&0.006	&1.507	&0.009	&2.022	&0.012	&2.353	&0.009	 &  2.055	&0.007	&1.522	&0.009	&2.028	&0.012	&2.293	&0.008	&54743.1\\
        &       &        &      &        &       &      &       &      1.808	&0.01	&1.301	&0.016	&1.771	&0.024	&2.149	&0.017	&55001.4\\
          &       &        &      &        &       &      &   &    1.811	&0.006	&1.296	&0.011	&1.872	&0.018	&2.19	&0.012	&55004.4\\
        &       &        &      &        &       &      &       &       1.816	&0.007	&1.332	&0.012	&1.807	&0.018	&2.188	&0.013	&55005.4\\
       &       &        &      &        &       &      &       &        1.82	&0.008	&1.326	&0.012	&1.82	&0.018	&2.168	&0.013	&55006.4\\
       &       &        &      &        &       &      &       &   1.81	&0.007	&1.303	&0.011	&1.811	&0.017	&2.206	&0.012	&55008.4\\
1.636	&0.006	&1.112	&0.006	&1.718	&0.008	&2.114	&0.008	 &  1.739	&0.005	&1.168	&0.006	&1.746	&0.008	&2.108	&0.007	&55068.2\\
1.611	&0.007	&1.101	&0.006	&1.703	&0.008	&2.106	&0.009	 &  1.72	&0.005	&1.175	&0.006	&1.722	&0.007	&2.123	&0.007	&55070.2\\
1.623	&0.005	&1.112	&0.006	&1.693	&0.008	&2.101	&0.007	 &  1.743	&0.005	&1.203	&0.006	&1.672	&0.007	&2.129	&0.006	&55071.2\\
1.609	&0.006	&1.099	&0.006	&1.712	&0.009	&2.079	&0.007	 & 1.699	&0.006	&1.201	&0.006	&1.675	&0.008	&2.117	&0.006	&55072.2\\
          &       &       &        &      &        &       &      &             1.734	&0.013	&1.169	&0.02	&1.709	&0.032	&2.175	&0.027	&55075.2\\
          &       &       &        &      &        &       &      &             1.724	&0.018	&1.111	&0.026	&1.76	&0.045	&2.248	&0.038	&55076.2\\
1.629	&0.007	&1.07	&0.008	&1.667	&0.013	&2.124	&0.009	 &  1.737	&0.007	&1.113	&0.007	&1.761	&0.011	&2.121	&0.008	&55087.2\\
1.641	&0.007	&1.067	&0.007	&1.645	&0.009	&2.106	&0.008	 &  1.74	&0.005	&1.144	&0.006	&1.69	&0.009	&2.109	&0.007	&55088.2\\
1.646	&0.007	&1.104	&0.007	&1.64	&0.011	&2.084	&0.008	 &  1.75	&0.007	&1.146	&0.007	&1.697	&0.011	&2.102	&0.009	&55089.2\\
1.65	&0.009	&1.097	&0.012	&1.622	&0.019	&2.085	&0.014	 &  1.737	&0.01	&1.15	&0.014	&1.68	&0.021	&2.147	&0.016	&55091.2\\
1.64	&0.005	&1.068	&0.006	&1.654	&0.008	&2.068	&0.007	 &  1.718	&0.005	&1.146	&0.007	&1.698	&0.009	&2.09	&0.007	&55092.2\\
1.857	&0.007	&1.139	&0.008	&1.788	&0.011	&2.187	&0.01	 &  1.912	&0.007	&1.218	&0.007	&1.758	&0.009	&2.193	&0.007	&55325.4\\
1.865	&0.007	&1.176	&0.008	&1.756	&0.012	&2.133	&0.009	 &  1.911	&0.007	&1.237	&0.009	&1.749	&0.012	&2.15	&0.01	&55334.4\\
1.872	&0.005	&1.166	&0.006	&1.784	&0.009	&2.152	&0.007	 &  1.923	&0.007	&1.256	&0.007	&1.769	&0.009	&2.159	&0.007	&55336.4\\
1.861	&0.007	&1.143	&0.009	&1.806	&0.014	&2.15	&0.011	 &   1.9	&0.008	&1.223	&0.009	&1.8	&0.014	&2.161	&0.011	&55339.4\\
1.897	&0.006	&1.284	&0.009	&1.733	&0.012	&2.106	&0.007	 &   1.971	&0.008	&1.281	&0.009	&1.769	&0.014	&2.148	&0.01	&55341.4\\
1.924	&0.007	&1.272	&0.007	&1.765	&0.01	&2.151	&0.006	 &   1.981	&0.008	&1.279	&0.008	&1.834	&0.01	&2.188	&0.008	&55352.4\\
1.932	&0.007	&1.237	&0.01	&1.861	&0.016	&2.234	&0.013	&   1.998	&0.008	&1.358	&0.01	&1.705	&0.012	&2.208	&0.011	&55354.4\\
1.95	&0.009	&1.283	&0.011	&1.798	&0.016	&2.174	&0.012	&   2.007	&0.006	&1.31	&0.009	&1.795	&0.013	&2.183	&0.01	&55358.3\\
1.939	&0.007	&1.262	&0.008	&1.813	&0.012	&2.18	&0.009	&   1.996	&0.006	&1.342	&0.008	&1.762	&0.011	&2.192	&0.008	&55362.3\\
1.946	&0.007	&1.251	&0.007	&1.859	&0.012	&2.154	&0.007	&   1.99	&0.006	&1.337	&0.008	&1.783	&0.011	&2.177	&0.008	&55365.3\\
1.94	&0.008	&1.296	&0.01	&1.83	&0.016	&2.195	&0.01	&   2.013	&0.012	&1.319	&0.013	&1.846	&0.019	&2.181	&0.015	&55368.4\\
1.891	&0.008	&1.321	&0.009	&1.872	&0.013	&2.176	&0.011	&   1.971	&0.009	&1.363	&0.008	&1.811	&0.011	&2.202	&0.01	&55378.4\\
1.9	&0.006	&1.293	&0.008	&1.837	&0.012	&2.218	&0.01	&   1.972	&0.007	&1.316	&0.009	&1.848	&0.012	&2.242	&0.01	&55380.3\\
1.933	&0.009	&1.325	&0.01	&1.868	&0.016	&2.248	&0.012	&   1.966	&0.006	&1.345	&0.01	&1.881	&0.016	&2.28	&0.01	&55384.4\\
1.879	&0.006	&1.342	&0.009	&1.798	&0.011	&2.209	&0.007	&   1.956	&0.006	&1.331	&0.009	&1.863	&0.012	&2.208	&0.01	&55386.4\\
         &       &       &        &      &        &       &           &1.992	&0.009	&1.393	&0.011	&1.82	&0.015	&2.216	&0.011	&55390.4\\
2.021	&0.011	&1.604	&0.014	&1.827	&0.016	&2.192	&0.012	 &   1.975	&0.006	&1.402	&0.008	&1.825	&0.011	&2.2	&0.009	&55392.3\\
1.84	&0.011	&1.425	&0.016	&1.863	&0.024	&2.314	&0.017	 &   1.939	&0.017	&1.402	&0.028	&1.909	&0.044	&2.213	&0.025	&55396.3\\
1.895	&0.011	&1.368	&0.014	&1.921	&0.022	&2.219	&0.018	 &   1.961	&0.009	&1.388	&0.01	&1.857	&0.014	&2.23	&0.012	&55399.3\\
1.84	&0.008	&1.363	&0.009	&1.864	&0.012	&2.248	&0.011	 &   1.955	&0.008	&1.365	&0.008	&1.887	&0.01	&2.252	&0.008	&55408.3\\
1.841	&0.005	&1.351	&0.009	&1.884	&0.014	&2.284	&0.01	 &   1.946	&0.007	&1.367	&0.008	&1.856	&0.012	&2.263	&0.009	&55410.3\\
1.847	&0.006	&1.343	&0.009	&1.881	&0.013	&2.296	&0.011	 &    1.947	&0.008	&1.384	&0.01	&1.89	&0.015	&2.293	&0.012	&55411.3\\
1.839	&0.02	&1.475	&0.028	&1.654	&0.035	&2.3	&0.036 &   1.906	&0.036	&1.329	&0.04	&1.919	&0.066	&2.279	&0.061	&55413.3\\
1.881	&0.007	&1.337	&0.011	&1.853	&0.017	&2.265	&0.012	 &   1.913	&0.013	&1.374	&0.019	&1.851	&0.028	&2.24	&0.024	&55418.3\\
1.884	&0.007	&1.358	&0.009	&1.814	&0.013	&2.263	&0.01	 &   1.978	&0.009	&1.359	&0.011	&1.874	&0.017	&2.251	&0.012	&55421.3\\
1.896	&0.008	&1.337	&0.009	&1.809	&0.013	&2.304	&0.011	 &   1.943	&0.008	&1.324	&0.01	&1.876	&0.015	&2.247	&0.011	&55425.3\\
1.945	&0.007	&1.352	&0.007	&1.909	&0.009	&2.276	&0.007	 &   1.949	&0.008	&1.366	&0.006	&1.844	&0.008	&2.26	&0.007	&55427.3\\
2.052	&0.008	&1.65	&0.012	&2.088	&0.017	&2.587	&0.015	&2.096	&0.008	&1.548	&0.01	&2.189	&0.017	&2.528	&0.014	&55735.4\\
2.026	&0.006	&1.601	&0.01	&2.156	&0.014	&2.584	&0.012	&2.105	&0.007	&1.603	&0.009	&2.127	&0.012	&2.499	&0.008	&55737.4\\
2.019	&0.008	&1.583	&0.01	&2.128	&0.015	&2.593	&0.013	&2.111	&0.006	&1.641	&0.009	&2.095	&0.012	&2.536	&0.011	&55739.4\\
2.083	&0.007	&1.492	&0.008	&2.073	&0.013	&2.549	&0.012	& 2.095	&0.008	&1.536	&0.008	&2.027	&0.011	&2.528	&0.011	&55762.3\\
2.059	&0.008	&1.499	&0.009	&2.11	&0.015	&2.576	&0.014	 &2.104	&0.009	&1.557	&0.01	&1.997	&0.014	&2.563	&0.014	&55763.2\\
2.101	&0.017	&1.49	&0.029	&2.018	&0.044	&2.563	&0.031	& & & & & & & & &55772.3\\
2.248	&0.007	&1.736	&0.012	&2.212	&0.016	&2.676	&0.013	&2.228	&0.01	&1.555	&0.012	&2.213	&0.02	&2.643	&0.018	&56164.8\\
2.17	&0.009	&1.637	&0.012	&2.23	&0.02	&2.668	&0.016	&2.253	&0.01	&1.582	&0.012	&2.366	&0.024	&2.623	&0.017	&56179.7\\
\hline
\end{longtable}
}

\clearpage
\fontsize{7}{9}\selectfont{\begin{longtable}{ccccccccc}

\caption{\label{datahv} V-and R-band photometry of HE\,0047-1756 , as in Fig.~\ref{fig:LChe0047V}.}\\

\hline
\hline
 mag $A_V$ & $\sigma_{A,V}$ & mag $B_V$ & ${\sigma}_{B,V}$&mag $A_R$ & $\sigma_{A,R}$ & mag $B_R$ & ${\sigma}_{B,R}$ &$MJD$\\
\hline 
\endfirsthead 
\caption{continued.}\\ 
 
\hline
\hline
  mag $A_V$ & $\sigma_{A,V}$ & mag $B_V$ & ${\sigma}_{B,V}$&mag $A_R$ & $\sigma_{A,R}$ & mag $B_R$ & ${\sigma}_{B,R}$ &$MJD$\\
\hline 
\endhead 
\hline
\endfoot
-0.21	&0.003	&1.251	&0.007	&-0.172	&0.003	&1.224	&0.007	&54624.4\\
-0.202	&0.003	&1.243	&0.008	&       &       &         &     &54626.4\\
-0.2	&0.002	&1.216	&0.005	&       &       &         &     &54628.4\\
-0.204	&0.002	&1.241	&0.006	&       &       &         &     &54633.4\\
-0.213	&0.004	&1.249	&0.011	&       &       &         &      &54640.4\\
-0.214	&0.004	&1.264	&0.009	&-0.194	&0.004	&1.229	&0.009	&54644.4\\
-0.221	&0.003	&1.251	&0.008	&-0.179	&0.004	&1.244	&0.008	&54646.4\\
-0.222	&0.003	&1.237	&0.007	&-0.195	&0.003	&1.239	&0.007	&54648.4\\
-0.23	&0.005	&1.222	&0.012	&-0.196	&0.005	&1.236	&0.012	&54650.4\\
-0.211	&0.002	&1.245	&0.005	&-0.195	&0.002	&1.249	&0.006	&54652.4\\
-0.216	&0.003	&1.252	&0.008	&-0.188	&0.002	&1.243	&0.006	&54654.4\\
-0.22	&0.003	&1.225	&0.006	&-0.213	&0.004	&1.233	&0.009	&54655.4\\
-0.229	&0.005	&1.227	&0.012	&-0.19	&0.004	&1.241	&0.01	&54656.4\\
-0.233	&0.004	&1.272	&0.011	&-0.199	&0.004	&1.259	&0.009	&54660.3\\
-0.236	&0.006	&1.194	&0.015	&-0.219	&0.004	&1.224	&0.009	&54662.3\\
 & & &                     &-0.202	&0.004	&1.248	&0.009	&54664.3\\
-0.246	&0.005	&1.174	&0.013	& & &  & &54670.3  \\                           
         -0.232	&0.005	&1.221	&0.011	&-0.203	&0.004	&1.204	&0.009	&54672.3\\
-0.256	&0.003	&1.17	&0.007	&-0.215	&0.004	&1.259	&0.009	&54674.3\\
-0.264	&0.003	&1.203	&0.008	&-0.207	&0.004	&1.204	&0.008	&54675.3\\
-0.26	&0.004	&1.193	&0.008	&-0.214	&0.006	&1.251	&0.013	&54677.3\\
-0.26	&0.004	&1.227	&0.01	&-0.218	&0.005	&1.22	&0.012	&54678.3\\
-0.275	&0.004	&1.19	&0.009	&-0.22	&0.004	&1.199	&0.009	&54681.3\\
-0.271	&0.004	&1.192	&0.009	&-0.218	&0.003	&1.192	&0.008	&54682.3\\
-0.264	&0.003	&1.17	&0.007	&-0.219	&0.004	&1.219	&0.009	&54684.3\\
-0.272	&0.005	&1.179	&0.011	& & & & &54686.3\\
 -0.27	&0.003	&1.183	&0.008	                              &-0.218	&0.004	&1.195	&0.008	&54688.3\\
 -0.272	&0.003	&1.201	&0.008            &-0.207	&0.005	&1.162	&0.011	&54690.3\\
-0.287	&0.004	&1.211	&0.011	&-0.219	&0.005	&1.168	&0.012	&54698.3\\
-0.257	&0.004	&1.183	&0.009	&       &       &         &&54700.3\\
-0.257	&0.004	&1.157	&0.009	&-0.217	&0.004	&1.156	&0.009	&54702.4\\
-0.265	&0.003	&1.173	&0.007	&-0.226	&0.003	&1.183	&0.008	&54704.2\\
-0.262	&0.004	&1.202	&0.008	&-0.198	&0.008	&1.188	&0.017	&54708.2\\
-0.269	&0.003	&1.201	&0.008	&-0.208	&0.005	&1.214	&0.011	&54710.2\\
-0.254	&0.004	&1.178	&0.01	&-0.215	&0.004	&1.161	&0.009	&54716.2\\
-0.262	&0.003	&1.189	&0.009	&-0.217	&0.005	&1.187	&0.011	&54720.3\\
        &       &       &         &            -0.226	&0.007	&1.181	&0.02	&54724.2\\
        &       &       &         &                                    -0.198	&0.005	&1.198	&0.013	&54726.2\\
          &       &       &         &                                  -0.201	&0.004	&1.17	&0.01	&54728.2\\
-0.257	&0.005	&1.162	&0.011	&       &       &         &&54730.3\\
  &       &       &         &                                              -0.23	&0.01	&1.258	&0.022	&54732.3\\
-0.266	&0.003	&1.176	&0.008	&   -0.214	&0.004	&1.169	&0.009	&54734.3\\
-0.269	&0.004	&1.172	&0.008	&     -0.216	&0.003	&1.155	&0.007	&54736.3\\ 
-0.265	&0.004	&1.161	&0.01	&       -0.215	&0.005	&1.183	&0.013	&54738.2\\ 
  &       &       &         &                                               -0.207	&0.004	&1.203	&0.008	&54740.2\\
  &       &       &         &                                               -0.344	&0.003	&1.092	&0.006	&55066.4\\

-0.455	&0.005	&1.03	&0.011	& -0.37	&0.005	&1.045	&0.011	&55067.4\\
-0.496	&0.006	&1.053	&0.015	&   -0.367	&0.005	&1.069	&0.011	&55068.3\\
-0.468	&0.005	&1.003	&0.011	&-0.361	&0.003	&1.098	&0.008	&55070.3\\
-0.451	&0.004	&1.01	&0.008	&-0.343	&0.003	&1.09	&0.008	&55071.4\\
-0.475	&0.004	&1.055	&0.01	&       &       &         &&55074.3\\
-0.478	&0.004	&0.997	&0.011	&-0.371	&0.003	&1.115	&0.008	&55075.3\\
-0.488	&0.006	&1.019	&0.019	&       &       &         &&55086.4\\
-0.517	&0.005	&1.007	&0.011	&-0.389	&0.003	&1.082	&0.008	&55087.3\\
-0.494	&0.005	&1.018	&0.011	&-0.394	&0.003	&1.074	&0.006	&55088.3\\
-0.488	&0.003	&0.971	&0.007	&-0.388	&0.003	&1.081	&0.007	&55089.4\\
-0.513	&0.004	&0.977	&0.01	&-0.395	&0.003	&1.068	&0.006	&55090.3\\
-0.514	&0.004	&0.996	&0.009	&-0.393	&0.003	&1.062	&0.006	&55092.3\\
-0.423	&0.01	&1.067	&0.026	&       &       &         &&55352.4\\
-0.41	&0.008	&1.069	&0.019	&-0.32	&0.003	&1.158	&0.008	&55355.4\\
-0.408	&0.007	&1.101	&0.017	&-0.325	&0.003	&1.177	&0.008	&55359.4\\
   &       &       &         &                                         -0.293	&0.005	&1.195	&0.013	&55367.4\\
-0.384	&0.006	&1.15	&0.017	&-0.28	&0.004	&1.189	&0.01	&55375.4\\
-0.376	&0.007	&1.124	&0.02	&-0.299	&0.004	&1.223	&0.011	&55377.4\\
-0.365	&0.01	&1.107	&0.025	&-0.283	&0.004	&1.169	&0.011	&55379.4\\
-0.361	&0.011	&1.139	&0.025	&-0.275	&0.004	&1.21	&0.01	&55387.4\\
      &       &       &         &                                       -0.284	&0.003	&1.222	&0.007	&55389.4\\
-0.346	&0.011	&1.163	&0.028	&-0.27	&0.003	&1.201	&0.008	&55391.4\\
                &       &       &         &                            -0.261	&0.003	&1.183	&0.009	&55393.4\\   
-0.323	&0.007	&1.133	&0.016	&-0.271	&0.004	&1.236	&0.01	&55397.3\\
-0.347	&0.008	&1.18	&0.021	&-0.269	&0.004	&1.226	&0.01	&55399.4\\
-0.327	&0.008	&1.149	&0.019	&-0.269	&0.005	&1.148	&0.011	&55400.4\\
-0.338	&0.006	&1.197	&0.017	&-0.262	&0.003	&1.223	&0.008	&55401.4\\
-0.349	&0.008	&1.202	&0.024	&-0.276	&0.004	&1.26	&0.011	&55403.4\\
-0.342	&0.007	&1.266	&0.019	&-0.251	&0.003	&1.205	&0.008	&55410.4\\
-0.339	&0.006	&1.225	&0.017	&-0.267	&0.004	&1.17	&0.01	&55414.4\\
-0.328	&0.007	&1.203	&0.019	&-0.267	&0.004	&1.255	&0.009	&55418.3\\
-0.325	&0.011	&1.22	&0.027	&-0.279	&0.004	&1.213	&0.01	&55427.3\\
    &       &       &         &                                      -0.314	&0.005	&1.206	&0.013	&55723.4\\                
   &       &       &         &                                        -0.353	&0.005	&1.212	&0.012	&55725.4\\
-0.422	&0.015	&1.154	&0.041	&       &       &         &&55726.4\\
-0.404	&0.013	&1.149	&0.034	&-0.336	&0.005	&1.211	&0.013	&55736.4\\
-0.425	&0.013	&1.163	&0.035	&       &       &         &&55737.4\\
-0.416	&0.009	&1.156	&0.023	&-0.349	&0.004	&1.214	&0.01	&55739.4\\
-0.4	&0.01	&1.217	&0.029	&-0.341	&0.005	&1.213	&0.015	&55760.4\\
-0.402	&0.011	&1.183	&0.032	&-0.328	&0.005	&1.24	&0.014	&55762.4\\
-0.414	&0.009	&1.225	&0.025	&-0.337	&0.004	&1.2	&0.011	&55765.4\\
-0.404	&0.012	&1.169	&0.03	&-0.347	&0.005	&1.235	&0.013	&55767.3\\
-0.398	&0.022	&1.281	&0.063	&-0.327	&0.006	&1.186	&0.014	&55769.4\\
-0.423	&0.009	&1.249	&0.023	&-0.348	&0.003	&1.255	&0.008	&55772.4\\
&&&& -0.366 &0.006 &1.203 &0.025&55775.4\\
-0.412	&0.01	&1.186	&0.027	&-0.353	&0.005	&1.249	&0.013	&55779.4\\
-0.404	&0.013	&1.136	&0.035	&-0.336	&0.006	&1.192	&0.015	&55784.4\\
-0.41	&0.009	&1.152	&0.024	&-0.358	&0.004	&1.258	&0.013	&55793.3\\
-0.402	&0.011	&1.184	&0.031	&-0.335	&0.005	&1.217	&0.015	&55794.3\\
-0.413	&0.009	&1.218	&0.026	&-0.347	&0.005	&1.231	&0.013	&55795.3\\
-0.404	&0.012	&1.17	&0.03	&-0.342	&0.006	&1.207	&0.014	&55804.3\\
-0.501	&0.011	&1.088	&0.029	&-0.405	&0.012	&1.107	&0.029	&56124.9\\
-0.513	&0.014	&1.099	&0.037	&-0.416	&0.01	&1.109	&0.025	&56127.9\\
-0.523	&0.008	&1.07	&0.026	&-0.448	&0.005	&1.178	&0.015	&56134.8\\
-0.525	&0.012	&1.069	&0.032	&-0.45	&0.009	&1.108	&0.022	&56135.9\\
-0.545	&0.01	&1.082	&0.03	&-0.454	&0.007	&1.154	&0.021	&56139.9\\
-0.533	&0.011	&1.021	&0.029	&       &       &         &&56164.8\\
-0.541	&0.01	&1.033	&0.027	&-0.467	&0.009	&1.096	&0.021	&56167.9\\
-0.532	&0.008	&1.117	&0.021	&-0.454	&0.006	&1.142	&0.017	&56179.8\\
-0.53	&0.012	&1.096	&0.031	&-0.435	&0.012	&1.089	&0.028	&56181.7\\
-0.535	&0.01	&1.11	&0.029	&-0.462	&0.009	&1.175	&0.025	&56186.7\\
\hline

\end{longtable}
}

\appendix

\section{}
\begin{table}[htbp]
\large
  \caption{Summary of the main input parameters for the PyCS spline (spl) and dispersion (disp) methods. Indices A and B refer to the quasar images. (See \cite{2013A&A...553A.120T} for detailed explanation.)}
\label{tab:PyCS}
\centering

\hskip-0.cm\begin{tabular}{p{3cm}p{2.5cm}}
\hline
\hline
$spl:{\eta}_{intr}$&$75 days$\\
$spl:{\eta}_{extr}$&$520 days$\\
$spl:\epsilon$&$15 days$\\
$spl:\alpha_{A}$&$1.1$\\
$spl:\beta_{A}$&$0$\\
$spl:\alpha_{B}$&$3.1$\\
$spl:\beta_{B}$&$0$\\
$disp:interpdist$&$10 days$\\
$disp:nparams$&$1$\\
$tsrand$&$5 days$\\
$truetsr$&$0 days$\\

\hline
\end{tabular}
\end{table}

\end{document}